\documentclass[onecolumn,draftcls]{IEEEtran}
\usepackage{graphicx,color}
\usepackage{amsmath}
\usepackage{latexsym}
\usepackage{amssymb}
\usepackage{comment}
\usepackage{cite}
\usepackage{url}
\usepackage[T3,T1]{fontenc}
\DeclareSymbolFont{tipa}{T3}{cmr}{m}{n}
\DeclareMathAccent{\invbreve}{\mathalpha}{tipa}{16}
\global\long\def\L{\mathsf{L}}
\global\long\def\D{\mathsf{D}}
\global\long\def\Sgen{\mathcal{S}_\mathsf{gen}}
\global\long\def\Kgen{\mathcal{K}_\mathsf{gen}}
\global\long\def\A{\mathcal{A}}

\newcommand{\VarX}{X_1}
\newcommand{\VarY}{X_2}

\newcommand{\calVarX}{{\cal X}_1}
\newcommand{\calVarY}{{\cal X}_2}

\newcommand{\varxun}{x_1^n}

\newcommand{\tX}{\overline{X}_1}
\newcommand{\tY}{\overline{X}_2}

\newcommand{\E}{\mbox{\bf E}}
\newcommand{\Var}{\mbox{\bf Var}}

\newcommand{\Dist}[1]{{#1}}

\begin{document}
\title{
Secrecy Amplification for Distributed Encrypted Sources 
with Correlated Keys using Affine Encoders
}
\author{Bagus Santoso and Yasutada Oohama}
\maketitle

\newtheorem{proposition}{Proposition}
\newtheorem{definition}{Definition}
\newtheorem{theorem}{Theorem}
\newtheorem{lemma}{Lemma}
\newtheorem{corollary}{Corollary}
\newtheorem{remark}{Remark}
\newtheorem{property}{Property}

\newcommand{\defeq}{:=}

\newcommand{\Qed}{\hbox{\rule[-2pt]{3pt}{6pt}}}
\newcommand{\beq}{\begin{equation}}
\newcommand{\eeq}{\end{equation}}
\newcommand{\beqa}{\begin{eqnarray}}
\newcommand{\eeqa}{\end{eqnarray}}
\newcommand{\beqno}{\begin{eqnarray*}}
\newcommand{\eeqno}{\end{eqnarray*}}
\newcommand{\ba}{\begin{array}}
\newcommand{\ea}{\end{array}}

\newcommand{\vc}[1]{\mbox{\boldmath $#1$}}
\newcommand{\lvc}[1]{\mbox{\scriptsize \boldmath $#1$}}
\newcommand{\svc}[1]{\mbox{\scriptsize\boldmath $#1$}}

\newcommand{\wh}{\widehat}
\newcommand{\wt}{\widetilde}
\newcommand{\ts}{\textstyle}
\newcommand{\ds}{\displaystyle}
\newcommand{\scs}{\scriptstyle}
\newcommand{\vep}{\varepsilon}
\newcommand{\rhp}{\rightharpoonup}
\newcommand{\cl}{\circ\!\!\!\!\!-}
\newcommand{\bcs}{\dot{\,}.\dot{\,}}
\newcommand{\eqv}{\Leftrightarrow}
\newcommand{\leqv}{\Longleftrightarrow}

\newcommand{\irr}[1]{{\color[named]{Red}#1\normalcolor}}		

\newcommand{\MEq}[1]{\stackrel{
{\rm (#1)}}{=}}

\newcommand{\MLeq}[1]{\stackrel{
{\rm (#1)}}{\leq}}

\newcommand{\ML}[1]{\stackrel{
{\rm (#1)}}{<}}

\newcommand{\MGeq}[1]{\stackrel{
{\rm (#1)}}{\geq}}

\newcommand{\MG}[1]{\stackrel{
{\rm (#1)}}{>}}

\newcommand{\MPreq}[1]{\stackrel{
{\rm (#1)}}{\preceq}}

\newcommand{\MSueq}[1]{\stackrel{
{\rm (#1)}}{\succeq}}

\newcommand{\MSubeq}[1]{\stackrel{
{\rm (#1)}}{\subseteq}}

\newcommand{\MSupeq}[1]{\stackrel{
{\rm (#1)}}{\supseteq}}

\newcommand{\vckone}{{\vc k}_1}
\newcommand{\vcktwo}{{\vc k}_2}
\newcommand{\vcxone}{{\vc x}_1}
\newcommand{\vcxtwo}{{\vc x}_2}
\newcommand{\vcyone}{{\vc y}_1}
\newcommand{\vcytwo}{{\vc y}_2}

\newcommand{\cvcxone}{\check{\vc x}_1}
\newcommand{\cvcxtwo}{\check{\vc x}_2}

\newcommand{\hvckone}{\widehat{\vc k}_1}
\newcommand{\hvcktwo}{\widehat{\vc k}_2}
\newcommand{\hvcxone}{\widehat{\vc x}_1}
\newcommand{\hvcxtwo}{\widehat{\vc x}_2}

\newcommand{\lvckone}{{\lvc k}_1}
\newcommand{\lvcktwo}{{\lvc k}_2}
\newcommand{\lvcxone}{{\lvc x}_1}
\newcommand{\lvcxtwo}{{\lvc x}_2}
\newcommand{\lvcyone}{{\lvc y}_1}
\newcommand{\lvcytwo}{{\lvc y}_2}

\newcommand{\clvcxone}{\check{\lvc x}_1}
\newcommand{\clvcxtwo}{\check{\lvc x}_2}

\newcommand{\hlvckone}{\widehat{\lvc k}_1}
\newcommand{\hlvcktwo}{\widehat{\lvc k}_2}
\newcommand{\hlvcxone}{\widehat{\lvc x}_1}
\newcommand{\hlvcxtwo}{\widehat{\lvc x}_2}
\newcommand{\rvccone}{{\vc C}_1}
\newcommand{\rvcctwo}{{\vc C}_2}
\newcommand{\rvckone}{{\vc K}_1}
\newcommand{\rvcktwo}{{\vc K}_2}
\newcommand{\rvcxone}{{\vc X}_1}
\newcommand{\rvcxtwo}{{\vc X}_2}
\newcommand{\rvcyone}{{\vc Y}_1}
\newcommand{\rvcytwo}{{\vc Y}_2}
\newcommand{\hrvcxone}{\widehat{\vc X}_1}
\newcommand{\hrvcxtwo}{\widehat{\vc X}_2}

\newcommand{\lrvckone}{{\lvc K}_1}
\newcommand{\lrvcktwo}{{\lvc K}_2}
\newcommand{\lrvcxone}{{\lvc X}_1}
\newcommand{\lrvcxtwo}{{\lvc X}_2}
\newcommand{\lrvcyone}{{\lvc Y}_1}
\newcommand{\lrvcytwo}{{\lvc Y}_2}
\newcommand{\rvcci}{{\vc C}_i}
\newcommand{\rvcki}{{\vc K}_i}
\newcommand{\rvcxi}{{\vc X}_i}
\newcommand{\rvcyi}{{\vc Y}_i}
\newcommand{\hrvcxi}{\widehat{\vc X}_i}
\newcommand{\vcki}{{\vc k}_i}
\newcommand{\vcsi}{{\vc s}_i}
\newcommand{\vcti}{{\vc t}_i}
\newcommand{\vcvi}{{\vc v}_i}
\newcommand{\vcwi}{{\vc w}_i}
\newcommand{\vcxi}{{\vc x}_i}
\newcommand{\vcyi}{{\vc y}_i}

\newcommand{\vcs}{{\vc s}}
\newcommand{\vct}{{\vc t}}
\newcommand{\vcv}{{\vc v}}
\newcommand{\vcw}{{\vc w}}
\newcommand{\vcx}{{\vc x}}
\newcommand{\vcy}{{\vc y}}

\begin{abstract}
This paper proposed the application of 
post-encryption-compression (PEC) to strengthen the secrecy in the case 
of distributed encryption where the encryption keys are correlated to 
each other. We derive the universal code construction for the 
compression and the rate region where codes with achievability and 
secrecy are obtainable.	

Our main technique is to use affine encoders which are constructed from 
certain linear encoders to encode the ciphertexts before sending them to 
public communication channels. We show that if the rates of linear codes 
are within a certain rate region:(1) 
information leakage on the original sources 
from the encoded ciphertexts without the keys 
is negligible, while (2) one who has 
legitimate keys is able to retrieve the original source data with 
negligible error probability.

\newcommand{\ZapSSS}{
In this paper, we consider
a system where
multiple sources are 
encrypted in separated nodes and sent through their respective
public communication channels into a joint sink node.
We are interested at  the problem on   
protecting the security 
of an already existing system such above,
which is found out to
have correlated encryption keys. 
In particular, we focus on
finding a solution 
without introducing additional secret keys and
with minimal modification 
to minimize the cost and
the risk of 
bringing down an already running system.
We propose a solution under
a security model where an eavesdropper obtains all 
ciphertexts, i.e., encrypted sources, 
by accessing available public communication channels.
Our main technique is to use affine encoders which are constructed from certain 
linear encoders to encode the ciphertexts before sending them to public communication channels.
We show that if the rates of linear codes are within a certain rate region:
(1) the success probability of any eavesdropper to extract
the original sources from the encoded ciphertexts  
without  the keys is negligible, while (2) one who 
has legitimate keys is able to retrieve the original 
source data with negligible error probability.
}
\end{abstract}


\section{Introduction \label{sec:introduction}}

\newcommand{\INtro}{
}{
\subsection*{Background}
In this paper, we consider the problem of strenghtening the security of  
communication in multi-source single-destination network.
Especially, we are interested on practical solutions with minimum
modifications which can be applied even on  already running systems.
More precisely, we consider a network system described
as follows: multiple sources $X_1$ and $X_2$ are processed 
in separated nodes, and then sent through their 
respective public communication channels
to a joint sink node. 
Now suppose that an already running system 
has a potential secrecy/privacy
problem such that  $(X_1,X_2)$ might be leaked to  
an adversary which is eavesdropping all public 
communication channel.

A common measure to prevent the leaking of 
$(X_1,X_2)$ to such eavesdropper 
is by encrypting each source using one time pad encryption in its respective corresponding node
before it is sent to the public channel.
For $i=1,2$, let $X_i$ be encrypted using key $K_i$ into $C_i=X_i\oplus K_i$. 
Instead of sending $X_1$ and $X_2$, the system sends
the ciphertexts $C_1$ and $C_2$ to public communication channels.
Obviously, if $K_1$ and $K_2$ are \emph{ideally} 
generated such that each is following uniform distribution and
is independent to each other, no problem is left as
$H(X_1X_2|C_1C_2)=H(X_1X_2)$ holds automatically. 
Note that this means that the pair of ciphertexts 
$(C_1,C_2)$ does not give any additional information 
about $(X_1,X_2)$ and thus
no one is able to reveal $(X_1,X_2)$ using $(C_1,C_2)$ 
with a better success
probability than that of randomly guessing $(X_1,X_2)$ 
based on the distribution of 
$(X_1,X_2)$ only (without knowing $(C_1,C_2)$).

\subsection*{Problem Framework: Secrecy/Privacy Guarantee under Correlated Keys in Distributed Encryption}

However, in real world, there is no guarantee that keys 
are always ideally generated,
and encryption keys in a system might be correlated to each other.
It is easy to see that if $K_1$ and $K_2$ are correlated to each other, 
i.e., $H(K_1|K_2)<H(K_1)$, the following automatically holds:
\begin{align}
	H(X_1X_2|C_1C_2)<H(X_1X_2).
\label{eq:corr_cipher}
\end{align}
Notice that the inequation (\ref{eq:corr_cipher}) 
means that we are posed with a security challenge.
The first reason is that (\ref{eq:corr_cipher}) means that
we can no longer directly guarantee that the 
ciphertexts in \emph{pair}, i.e., $(C_1,C_2)$,
do not give additional information about $(X_1,X_2)$ to the 
eavesdropper.
Furthermore, (\ref{eq:corr_cipher}) still holds
although $K_1$ and $K_2$ are generated randomly 
from uniform distribution 
over their respective domain 
such that any \emph{single} separated ciphertext do not reveal
additional information about the corresponding source data, i.e.,
$H(X_i|C_i)=H(X_i)$ for $i=1,2$. 
In other words, we \emph{lost the security guarantee for secrecy} against eavesdroppers which
access all public communication channels when key are correlated to each other.

Here, we restate our research problem into the following question:
\textit{%
	Is there any method which: (1) strengthens the secrecy 
	such that it can guarantee that under the condition shown by inequation
	(\ref{eq:corr_cipher}) the eavesdropper can not easily extract $(X_1,X_2)$ from $(C_1,C_2)$,
	and (2) is implementable with small cost even on already 
	running systems? 	
}


\subsection*{Search for Solution}%
Since the root of the problem posed by (\ref{eq:corr_cipher}) is the correlation between the keys $K_1$ and $K_2$, 
it is natural to think that if we can somehow reduce the effect of 
correlation between $K_1$ and $K_2$ 
in $(C_1,C_2)$, then we might be able to amplify the secrecy to an extent that we can 
guarantee a certain level of secrecy close to perfect secrecy.

\ \par \noindent{}\textit{Na\"{\i}ve method 
(Additional New Secret Randomness):\ }
A na\"{\i}ve method to reduce the correlation between $K_1$ and $K_2$ is 
by introducing additional independent randomness to each node. 
For example, we can put additional independently generated
randomness $R_1$  and $R_2$ to each node
respectively, and use $K_1\oplus R_1$ and $K_2\oplus R_2$ as the 
new inputs as keys to the encryption. 
However, this method has serious drawbacks. Firstly, it requires
new private channels to send the randomness and secondly,
it requires 
each node to bear additional security costs
that each node has additional private storage to keep the new randomness securely
in private manner.  
We conclude that this na\"{\i}ve method is not feasible to implement in real world, especially
for an already running system and
for a system with nodes of lightweight devices such as 
wireless sensors.

\ \par

Since a simple additional secret randomness technique as above
sounds impractical in real world,
we need to find approach to reduce the effect of correlation
between $K_1$ and $K_2$ in $(C_1,C_2)$  without new secret randomness.

\subsection*{Fundamental Idea for Practical Solution: Compression of Keys}%
%
Our fundamental idea for solution is based on
our intuition that  \textit{the correlation between compressed keys is 
smaller than correlation of uncompressed keys}.
We can explain our intuition as follows. First, recall  
that the amount of correlation between two random sources $K_1,K_2$
is directly proportional to the mutual information between 
$K_1$ and $K_2$, i.e., $I(K_1;K_2)$.\footnotemark
For simplicity, let $K_1$ and $K_2$ are taking values from 
the same set of $n$ dimensional vectors
$\mathcal{K}^{n}$.
Let $\varphi$ be a mapping from $\mathcal{K}^{n}$  onto a set of $m$ dimensional
vectors $\mathcal{K}^{m}$, where
where $n>m$. One may simply treat $\varphi$ as a kind of compression function. 
Using the fact that $H(K_2|\varphi(K_1))\geq H(K_2|K_1)$ and $H(\varphi(K_1)|\varphi(K_2))\geq H(\varphi (K_1)|K_2)$ hold,
it is easy to derive the following inequations.
\footnotetext{%
	Also recall that when $K_1$ and $K_2$ are independent and  have no correlation, 
	$I(K_1;K_2)=0$ holds,
	while if $K_1$ and $K_2$ are not independent and have some correlation, $I(K_1;K_2)>0$.
}

\begin{align}
	I(K_1;K_2)\geq I(\varphi(K_1),K_2)\geq I(\varphi(K_1);\varphi (K_2)).	\label{eq:compression}
\end{align}

The above inequations (\ref{eq:compression}) basically says that compressing the keys $(K_1,K_2)$ may 
reduce the effect of correlation between them. 
Thus, one immediate  approach is to
compress the keys \emph{directly} before inputting them into the encryption process.

\ \par \noindent{}\textit{Infeasibility of Direct Compression of Keys:\ }
However, recall that  there are two points of inputs
to the encryption in each node, i.e., \emph{source} and \emph{key},
Thus, if we want to use the compressed keys as the new keys to the encryption,
in order to guarantee secrecy, in general, 
we also have to compress the messages. 
Especially in the case of one-time pad encryption, 
we need to compress the messages to an extent that the lengths are same with the compressed keys.
This means we have to perform compression \emph{two times} for each node.
Moreover, in real world, the devices at the nodes might have
the keys \emph{hardwired} into the 
electronic circuit, and thus modification
of the keys before the encryption will require us a special technique
to perform a hardware intrusion without bringing down
the already running system.
Obviously, this kind of modification is  risky or impossible in
some cases.
Therefore, we conclude that direct compression of keys at the point of 
inputs to encryption is practically infeasible  in general.

\ \par

Hence, we narrow the research question into the following: \textit{Can we find a better 
method for compression such that we do not need to modify inputs of encryption directly
and requires  less number of compressions than two times for each node?}

\subsection*{Proposed Solution: Indirect Compression of Keys through Compression of Ciphertexts using Affine Encoders}%


The main result in this paper is that 
we discover a method to perform compression on the keys \emph{indirectly} by
compressing the ciphertexts. 
We only
need to perform the compression \emph{only once} for each node and thus the implementation
cost is only about \emph{half} of the method which performs compressions on inputs before encryption
described above.
The core of our discovery is the specific
construction of an \emph{affine encoder} 
as a \emph{good} compression function. 
We prove that the result of compression of a ciphertext 
from one-time pad encryption
using our affine encoder 
can be seen as 
one-time pad encryption of a \emph{compressed message} with 
a \emph{compressed key}.
%
%

As a illustration, for $i$-th node ($i=1,2$),
let \emph{affine encoder} $\varphi_i$ be 
associated with a \emph{linear encoder} $\phi_i$  and a vector $a_i$,
and  let $\varphi_i$ be defined such that 
$\varphi_i(x)=\phi_i(x)\oplus a_i$.
Using $\varphi_i$, we compress the ciphertext of $i$-th node, 
$C_i=X_i \oplus K_i$, into 
$\widetilde{C}_i=\varphi_i(X_i\oplus K_i)
=\phi(X_i\oplus K_i)\oplus a_i$.
Thanks to the \emph{homomorphic} property of linear encoder $\phi_i$,
we can expand $\phi_i(X_i\oplus K_i)$ into $\phi(X_i)\oplus \phi_i(K_i)$ and
we obtain the following equation. 
\begin{align}
\widetilde{C}_i
=\varphi_i(C_i)
=\varphi_i(X_i\oplus K_i)
=\phi_i(X_i)\oplus \phi_i(K_i) \oplus a_i
=\phi_i(X_i)\oplus \varphi_i(K_i)
=\phi_i(X_i)\oplus \widetilde{K}_i.
\label{eq:affine_expand_intro}
\end{align}
Here we set $\widetilde{K}_i \defeq \varphi_i(K_i),i=1,2$.
We can see $\phi_i(X_i)$ as the compressed source and 
$\widetilde{K}_i$
as the compressed key. Hence, an eavesdropper 
which collects $(\widetilde{C}_1,\widetilde{C}_2)$ 
from public communication channels
will see $(Y_1,Y_2)$ as results of one-time pad encryption with
compressed keys $(\varphi_1(K_1),\varphi_2(K_2))$ which has
less correlation compared to the original keys $(K_1,K_2)$.


We borrow the technique of Oohama\cite{Oohama:2007:IRP:1521152.1521168} on
generating randomness using Slepian-Wolf coding \cite{1055037}
to make the joint distribution of compressed keys which
are hidden within the compressed ciphertexts 
\emph{exponentially} close to the uniform distribution that
the effect of correlation between the keys becomes negligible.
Furthermore, we  borrow the result of Csisz\'{a}r
\cite{Csiszr1982LinearCF} to show that we can obtain good linear encoders and decoders
such that in joint sink node 
the original sources data can be retrieved
with \emph{exponentially} negligible error probability.

We prove that the code construction can be made to depend on only 
transmission rates using the universal code technique.
As far as our knowledge, our result is the first to show \emph{explicitly}
that the \emph{preserving of code structure} which is the property of 
affine encoders constructed from linear encoders
is \emph{essential} in order to amplify the secrecy
and to preserve the achievability at the same time in the case
of distributed encodings/encryption.
One can see that our result is in parallel with the existing work of 
K{\"o}rner and Marton \cite{KornerMarton} in the sense that
\cite{KornerMarton} 
shows that the \emph{preserving of code structure} by
linear encoders is \emph{essential}
in order to prove the optimal transmission rate in the case
of two helper network systems.

\subsection*{Practical Feasibility of Proposed Solution}
In practice, our approach does not require
hardware intrusion to the terminal devices.
We can modify the output of the encryption easily by simply
connecting the already existing device in each node with  
an additional external equipment which is capable to receive the ciphertext
from the encryption process as inputs, encode
them using specified linear codes, and then
finally output the encoded ciphertext to the public communication channel.
In order to prevent that the leak of pre-encoded original ciphertexts to
the eavesdropper in the case of wireless communication, 
we can apply a simple idea to enclose the existing device 
and the additional equipment in 
a Faraday cage so that no electronic signal carrying the 
pre-encoded ciphertexts
leaks outside. 

\noindent{}\textit{On modification of joint sink node:\ }
We  remark that our proposed solution which will be described in detail
at later sections actually requires the modification of
the input and the output of the joint sink node. We argue that 
despite this requirement, our approach is still feasible and practical.
We can consider the joint sink node as a kind of 
information processing center in real world. 
And it is quite natural to assume that in such center,
the processing tasks are carried by
general-purpose machines with high modularity, that
the components are easy to be separated, modified, 
and recombined without disrupting the already running system. 

\subsection*{Related Works}
Randomness generation problem in distributed networks with multi-terminals 
has been researched by Muramatsu et al. \cite{Muramatsu03}, 
Oohama \cite{Oohama:2007:IRP:1521152.1521168}. However, these works only focus
on the secrecy and randomness issue and do not take into account the issue of  
achievability. Csisz\'{a}r has shown in \cite{Csiszr1982LinearCF} that
one can easily use linear codes to construct universal coding 
for all achievable rates in Slepian-Wolf networks. However, this work only focus
on the achievability and do not touch security related issues.

Johnson et al. \cite{1337277} has proposed a model similar to our setting in a sense
that they try to reach both achievability and secrecy at the same time using a similar
encryption-then-compress paradigm. However, they only focused on achieving 
achievability and secrecy using  a specific kind of encoders and compression method and
they do not show whether the encoders and compression method satisfy the universality.
In this paper, we show a more general results in the sense that we show that \emph{any} 
$\emph{good}$ linear encoder which attain random coding error exponent  can be used
to construct compression function which satisfy both achievability and secrecy and also
we show that our construction satisfies the universality.
Moreover, Johnson et al \cite{1337277} only consider secrecy in \emph{asymptotic} setting,
while in this paper we consider the secrecy in  \emph{concrete} setting with concrete exponential 
upper bound of the success probability of an eavesdropper revealing the sources from 
compressed ciphertexts in public channel.
}

\section{Preliminaries}
In this section, we show the basic notations and related consensus used in this paper. Also, we
explain the basic system setting and basic adversarial model we consider in this paper.

\ \par \noindent{}\textit{Random Sources of Information and Keys: \ }
Let $(X_1,X_2)$ be a pair of random variables from a finite set 
$\mathcal{X}_1\times \mathcal{X}_2$. 
Let $\{(X_{1,t},X_{2,t})\}_{t=1}^\infty$ be a stationary discrete
memoryless source(DMS) such that for each $t=1,2,\ldots$, the pair $(X_{1,t},X_{2,t})$ takes 
values in finite set $\mathcal{X}_1\times \mathcal{X}_2$ and obeys the same distribution 
as that of $(X_1,X_2)$ denoted by 
$P_{X_1 X_2}=\{P_{X_1 X_2} (x_1,x_2)\}_{(x_1,x_2)\in\mathcal{X}_1\times \mathcal{X}_2}$.
The stationary DMS $\{(X_{1,t},X_{2,t})\}_{t=1}^\infty$ 
is specified with $P_{X_1X_2}$.
Also, let $(K_1, K_2)$ be pair of random variables taken from 
the same  finite set $\mathcal{X}_1\times \mathcal{X}_2$ 
representing the pair of keys used for encryption at two separate terminals, of 
which the detailed description will be presented later.
Similarly, let $\{(K_{1,t},K_{2,t})\}_{t=1}^\infty$ be a stationary discrete
memoryless source such that for each $t=1,2,\ldots$, the pair $(K_{1,t},X_{K,t})$ takes values in finite set
$\mathcal{X}_1\times \mathcal{X}_2$ and obeys the same distribution as that of $(K_1,K_2)$ denoted by 
$P_{K_1 K_2}=\{P_{K_1 K_2} (k_1,k_2)\}_{(k_1,k_2)\in\mathcal{X}_1\times \mathcal{X}_2}$.
The stationary DMS $\{(K_{1,t},K_{2,t})\}_{t=1}^\infty$ 
is specified with $P_{K_1K_2}$.

\ \par\noindent{}\textit{Random Variables and Sequences: \ }
We write the sequence of random variables with length $n$ 
from the information source as follows:
${\rvcxone}\defeq X_{1,1}X_{1,2}\cdots X_{1,n}$, 
${\rvcxtwo}\defeq X_{2,1}X_{2,2}\cdots X_{2,n}$.
Similarly, the strings with length $n$ of $\mathcal{X}_1^n$ and $\mathcal{X}_2^n$ 
are written as ${\vcxone}\defeq x_{1,1}x_{1,2}\cdots
x_{1,n}\in\mathcal{X}_1^n$ and ${\vcxtwo}
\defeq
x_{2,1}x_{2,2}\cdots x_{2,n}\in\mathcal{X}_2^n$ respectively.
For $({\vcxone}, {\vcxtwo})\in\mathcal{X}_1^n\times \mathcal{X}_2^n$, 
$P_{{\lrvcxone}{\lrvcxtwo}}({\vcxone},{\vcxtwo})$ stands for the 
probability of the occurrence of $({\vcxone}, {\vcxtwo})$. 
When the information source is memoryless 
specified with $P_{X_1X_2}$, we have the following equation holds:
$P_{{\lrvcxone} {\lrvcxtwo}}({\vcxone},{\vcxtwo})=
\prod_{t=1}^n P_{X_1X_2}(x_{1,t},x_{2,t})$.
In this case we write $P_{{\lrvcxone} {\lrvcxtwo}}({\vcxone},{\vcxtwo})$
as $P_{X_1 X_2}^n({\vcxone},{\vcxtwo})$.
Similar notations are used for other random variables and sequences.

\ \par\noindent{}\emph{Consensus and Notations: }
Without loss of generality, throughout this paper,
we assume that $\mathcal{X}_1$ and $\mathcal{X}_2$ are finite fields.
The notation $\oplus$ is used to denote the field addition operation,
while the notation $\ominus$ is used to denote the field subtraction 
operation, i.e., $a\ominus b = a \oplus (-b)$ for any
elements $a,b$ of a same finite field. All discussions and 
theorems in this paper still hold although
$\mathcal{X}_1$  and $\mathcal{X}_2$  are different finite
fields. However, for the sake of simplicity, we use the same
notation for field addition and subtraction for both $\mathcal{X}_1$
and $\mathcal{X}_2$.


\begin{figure}[t]
\centering
\includegraphics[width=0.60\textwidth]{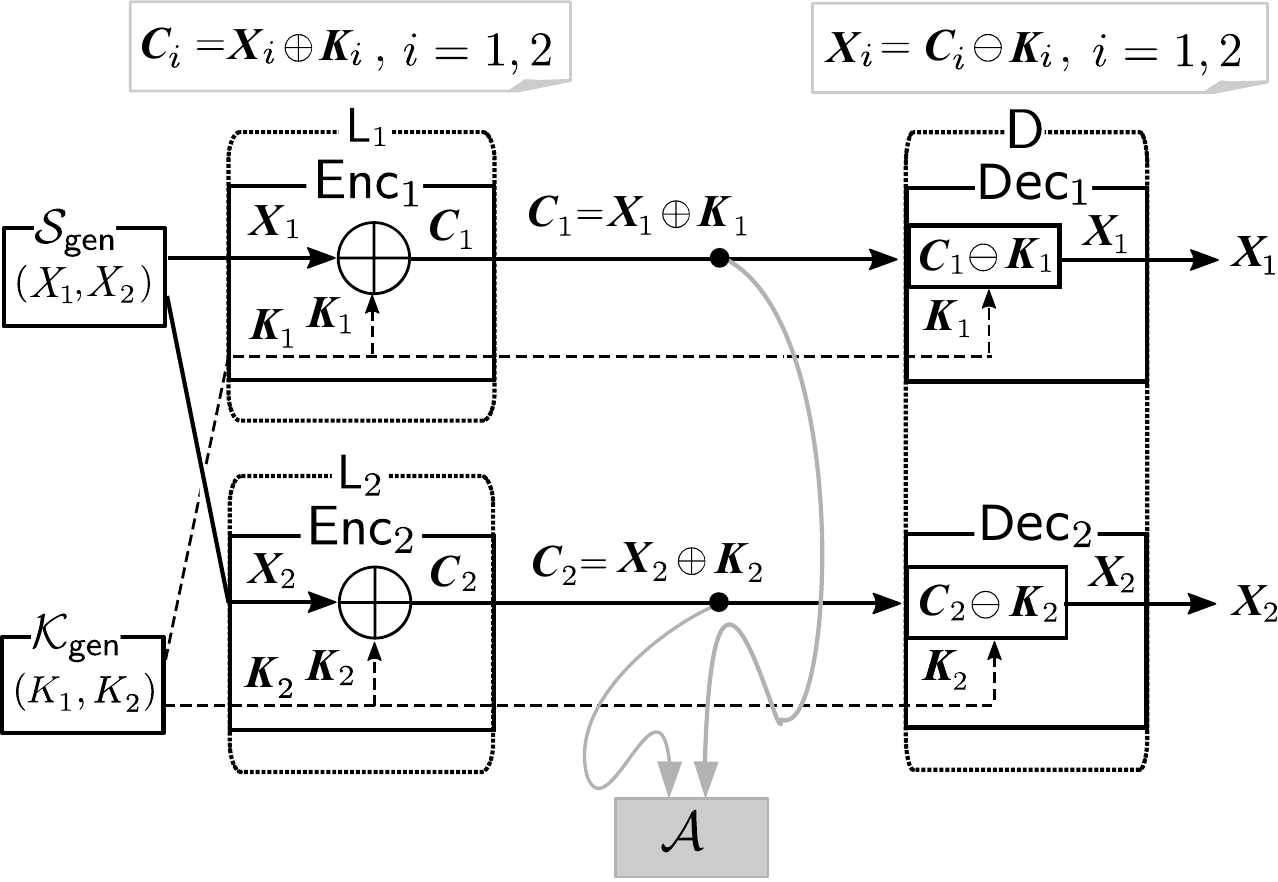}
\caption{Separate encryption of two corelated sources with 
joint decryption.
\label{fig:main}}
\end{figure}


\subsection{Basic System Description}

First, let the information sources and keys be generated independently by different parties
$\Sgen$ and $\Kgen$ respectively.
In our setting, we assume the followings.
\begin{itemize}
	\item 	The random keys ${\rvckone}$  and ${\rvcktwo}$ are generated by $\Kgen$
		from uniform distribution.
	\item	The key ${\rvckone}$ is correlated to  
		${\rvcktwo}$. 
	\item  	The sources ${\rvcxone}$ and ${\rvcxtwo}$ are generated by $\Sgen$ and are correlated
		to each other.
	\item   The sources are independent to the keys.
\end{itemize}
Next, let the two correlated 
random sources ${\rvcxone}$ and ${\rvcxtwo}$ from $\Sgen$
be sent to two separated nodes $\mathsf{L}_1$ and $\mathsf{L}_2$
respectively.
And let two random key (sources) ${\rvckone}$ and ${\rvcktwo}$ from $\Kgen$
be also sent separately to 
$\mathsf{L}_1$ and $\mathsf{L}_2$.
Further settings of our system are described as follows, 
as shown in Fig. \ref{fig:main}.
\begin{enumerate}
	\item \emph{Distributed Sources Processing:}
	At node $\L_1$, ${\rvcxone}$ is encrypted with the key ${\rvckone}$ using 
	encryption scheme $\mathsf{Enc}_1$, and at node $\L_2$,
	${\rvcxtwo}$ is encrypted with the key ${\rvcktwo}$ 
	using  encryption scheme $\mathsf{Enc}_2$. 
        The ciphertexts ${\rvcci},i=1,2$ are defined by
        $$
        {\rvcci} \defeq \mathsf{Enc_i}({\rvcxi},{\rvcki})
	={\rvcxi}\oplus {\rvcki}.
        $$

	\item \emph{Transmission:}
	Next, the ciphertexts ${\rvccone}$ and ${\rvcctwo}$ are sent to a common 
        information processing center $\D$ through two separated 
	\emph{public} communication channels.
	Meanwhile, the keys ${\rvckone}$ and ${\rvcktwo}$  are sent to $\D$
	through \emph{private} communication channels.

	\item \emph{Joint Sink Node Processing:}
	In $\D$, we decrypt the ciphertexts 
        $({\rvccone}, {\rvcctwo})$ using the keys 
        $({\rvckone},{\rvcktwo})$
	through the corresponding decryption procedure 
        $\mathsf{Dec}_i,i=1,2$ which is 
        defined by 
	$\mathsf{Dec}_i({\rvcci}, {\rvcki})=({\rvcci} \ominus {\rvcki})$ 
        for $i=1,2$. It is obvious that for each $i=1,2$,
        we can correctly reproduce the source outputs $\rvcxi$ 
        from $\rvcci$ and $\rvcki$ 
        by the decription function $\mathsf{Dec}_i$.
\end{enumerate}

\subsection*{Eavesdropper Adversarial Model (Informal Description)}
An \emph{eavesdropper adversary} $\A$ eavesdrops all public 
communication channels in the system and output/estimate
the original data from information sources.



%
\section{Proposed Idea: Affine Encoders as Privacy Amplifier}
%
Let $\phi^{(n)}:=(\phi_1^{(n)},\phi_2^{(n)})$ be 
a pair of linear mappings 
$\phi_1^{(n)}:\mathcal{X}_1^n\rightarrow \mathcal{X}_1^{m_1}$
and 
$\phi_2^{(n)}:\mathcal{X}_2^n\rightarrow \mathcal{X}_2^{m_2}$.
For each $i=1,2$, we define 
the mapping $\phi_i^{(n)} {\cal X}_i^n \to {\cal X}_i^{m_i}$
by 
\beq
\phi_i^{(n)}({\vcxi})
={\vcxi} A_i \mbox{ for }{\vcxi} \in {\cal X}_i^n,
\label{eq:homomorphica}
\eeq
where $A_i$ is a matrix with $n$ rows 
and $m_i$ columns. For each $i=1,2$, entries of $A_i$ are 
from ${\cal X}_i$. We fix $b_i^{m_i}\in \mathcal{X}_i^{m_i},$ 
$i=1,2$. For each $i=1,2$, define the mapping 
$\varphi^{(n)}_{i}: {\cal X}_i^n \to {\cal X}_i^{m_i}$
by 
\begin{align}
&\varphi_{i}^{(n)}({\vcki}):=
\phi_i^{(n)}({\vcki})\oplus b_i^{m_i}
={\vcki}A_i\oplus b_i^{m_i}, 
\notag\\
&\quad \mbox{ for }{\vcki} \in \mathcal{X}^n_i.
\label{eq:homomorphic}
\end{align}
For each $i=1,2$, the mapping $\varphi_{i}^{(n)}$ 
is called the affine mapping induced by the linear mapping 
$\phi_{i}^{(n)}$ and constant vector $b_i^{m_i}$ $\in{\cal X}^{m_i}$.
By the definition (\ref{eq:homomorphic}) of 
$\varphi_i^{(n)}$, $i=1,2$, those satisfy the following 
affine structure: 
\begin{align}
&\varphi_i^{(n)}({\vcxi} \oplus {\vcki})=
({\vcxi} \oplus {\vcki})A_i\oplus b_i^{m_i}
\notag\\
&={\vcxi} A_i\oplus({\vcki}A_i\oplus b_i^{m_i})
=\phi_i^{(n)}({\vcxi})\oplus \varphi_i^{(n)}({\vcki}),
\notag\\
&\quad \mbox{ for } {\vcxi}, {\vcki} \in {\cal X}_i^n.
\label{eq:affine}
\end{align}
Set $\varphi^{(n)}:=(\varphi_1^{(n)},\varphi_2^{(n)})$. 
Next, let $\psi^{(n)}$ be the corresponding joint decoder 
for $\phi^{(n)}$ such that
$
\psi^{(n)}:\mathcal{X}_1^{m_1} 
\times \mathcal{X}_2^{m_2} \rightarrow 
\mathcal{X}_1^{n} \times \mathcal{X}_2^{n}.
$
Note that $\psi^{(n)}$ does not have a linear structure in general. 
\subsubsection*{%
Description of Proposed procedure
}
We describe the procedure of our privacy amplified system as follows. 

\begin{figure*}[t] 
	\centering 
	\includegraphics[width=0.80\textwidth]{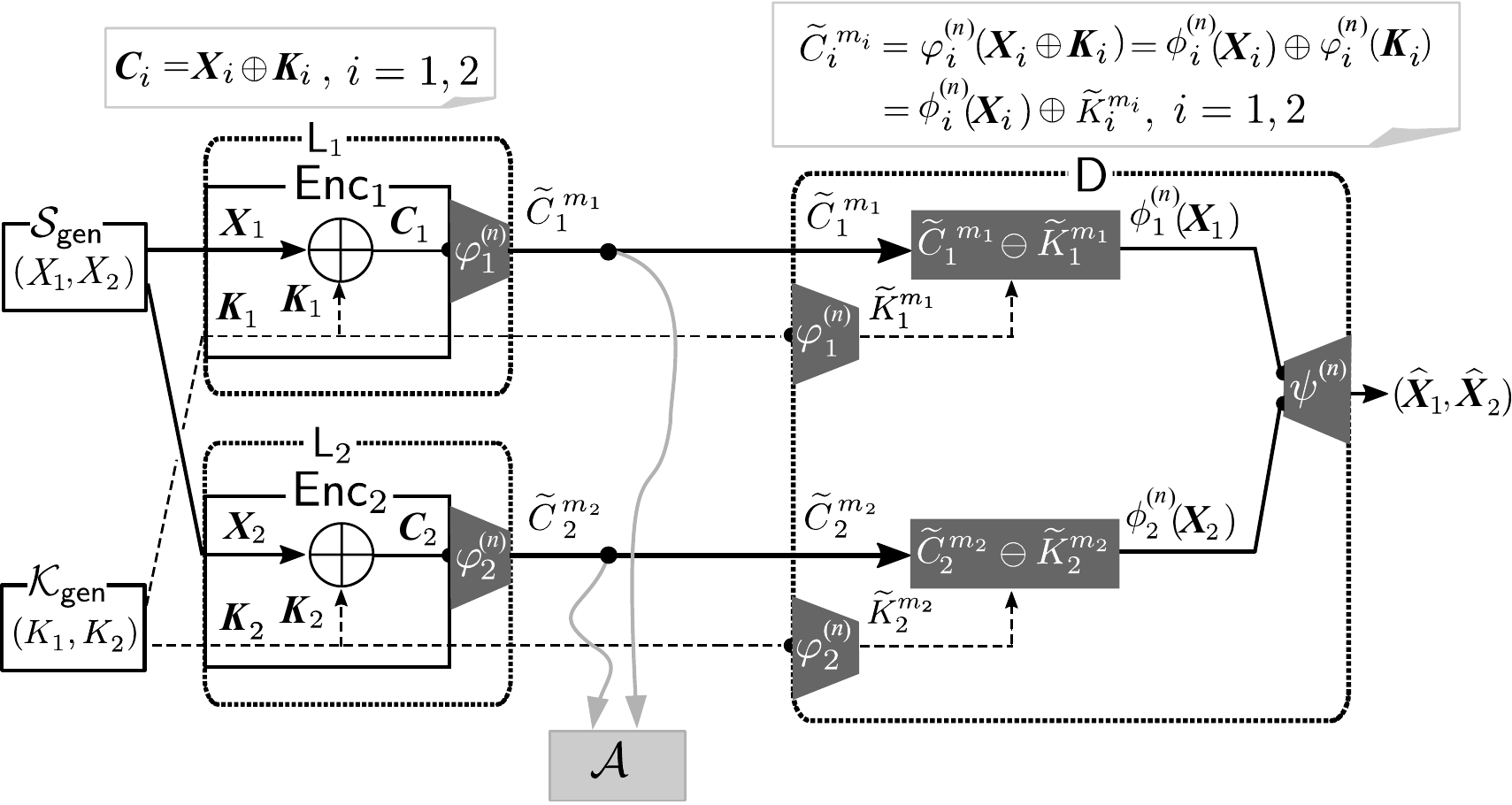}
	\caption{Our proposed solution: linear encoders 
	as privacy amplifiers.
\label{fig:solution}}%
\end{figure*}

\begin{enumerate}
	\item\emph{Encoding of Ciphertexts:}
	First, we use $\varphi_1^{(n)}$ and $\varphi_2^{(n)}$
	to encode the ciphertexts $C_1^{n}=X_1^{n}\oplus K_1^{n}$
	and $C_2^{n}=X_2^{n}\oplus K_2^{n}$.
	Let $\widetilde{C}_i^{m_i}=\varphi_i^{(n)}({\rvcci})$ for $i=1,2$.
	Then, instead of sending ${\rvccone}$ and ${\rvcctwo}$, 
	we send $\tilde{C}_1^{m_1}$ and $\tilde{C}_2^{m_2}$
	to public communication channel.
        By the affine structure (\ref{eq:affine})
        of encoders we have that for each $i=1,2$,  
        \begin{align}
        &\widetilde{C}_i^{m_i}=\varphi_i^{(n)}({\rvcxi}\oplus {\rvcki})
        =\phi_i^{(n)}({\rvcxi}) \oplus \varphi_i^{(n)}({\rvcki})
        \notag\\
        &=\widetilde{X}_i^{m_i} \oplus \widetilde{K}_i^{m_i},
        \label{eqn:aSdzx} 
        \end{align}
        where 
        $\widetilde{X}_i^{m_i} \defeq \phi_i^{(n)}({\rvcxi}),
        \widetilde{K}_i^{m_i} \defeq \varphi_i^{(n)}({\rvcki}).$ 
	\item\emph{Decoding at Joint Sink Node $\D$:}
	First, using the pair of linear encoders
	$(\varphi_1^{(n)},\varphi_2^{(n)})$,
	$\D$ encodes the keys $({\rvckone},{\rvcktwo})$ which are received
	through private channel into
	$(\widetilde{K}_1^{m_1},\widetilde{K}_2^{m_2})=$
	$(\varphi_1^{(n)}({\rvckone}),\varphi_2^{(n)}({\rvcktwo}))$.
	Receiving $(\widetilde{C}_1^{m_1},\widetilde{C}_2^{m_2})$ from
	public communication channel, $\D$ computes
	$\widetilde{X}_i^{m_i},i=1,2$ in the following way.
        From (\ref{eqn:aSdzx}), we have that for each $i=1,2$, 
        the decoder $\D$ can obtain 
        $\widetilde{X}_i^{m_i} = \phi_i^{(n)}({\rvcxi})$
        by subtracting 
        $\widetilde{K}_i^{m_i}=\varphi_i^{(n)}({\rvcki})$ 
        from $\widetilde{C}_i^{m_i}$. 
	Finally, $\D$ outputs $(\hrvcxone, \hrvcxtwo)$
	by applying the joint decoder 
        $\psi^{(n)}$ to 
	$(\widetilde{X}_1^{m_1},\widetilde{X}_2^{m_2})$ as follows:
	\begin{align}
		(\hrvcxone, \hrvcxtwo)
		&=(\psi^{(n)}(\widetilde{X}_1^{m_1},
                  \widetilde{X}_2^{m_2})) 
                \notag\\
	 	&=(\psi^{(n)} (\phi_1^{(n)}({\rvcxone}),
                              \phi_2^{(n)}({\rvcxtwo})). 
                 \label{eq:source_estimation}
	\end{align}
\end{enumerate}
Our privacy amplified system described above is illustrated 
in Fig. \ref{fig:solution}.

\subsection*{On Reliability}
From (\ref{eq:source_estimation}), 
it is clear that the decoding error probability $p_{\rm e}$ is as follows:
\begin{align*}
&p_{\rm e}=p_{\rm e}(\phi^{(n)},\psi^{(n)}|P_{X_1X_2}^n)
\\
&:= \Pr[\psi^{(n)}(\phi_1^{(n)}({\rvcxone}),\phi_2^{(n)}({\rvcxtwo}))\neq 
({\rvcxone},{\rvcxtwo})]. 
\end{align*}

\subsection*{On Security}
An eavesdropper $\A$ tries to estimate 
$({\rvcxone},{\rvcxtwo})\in \mathcal{X}_1^n \times \mathcal{X}_2^n$ from
$
(\widetilde{C}_1^{m_1},\widetilde{C}_2^{m_2})
=(\varphi_1^{(n)}({\rvcxone}\oplus {\rvckone}),
\varphi_2^{(n)}({\rvcxtwo}\oplus {\rvcktwo}))
\in\mathcal{X}_1^{m_1}\times\mathcal{X}_2^{m_2}
$. 
The information leakage $\Delta^{(n)}$ on $(\rvcxone,\rvcxtwo)$ 
from $(\widetilde{C}_1^{m_1},\widetilde{C}_2^{m_2})$
is measured by the mutual information 
between those two random pairs. This quantity is formally defined by 
\begin{align*}
\Delta^{(n)}=&\Delta^{(n)}(\varphi^{(n)}|{P}_{X_1X_2}^n,{P}_{K_1K_2}^n)
\\
\defeq & I(\widetilde{C}_1^{m_1},\widetilde{C}_2^{m_2};\rvcxone,\rvcxtwo).  
\end{align*}

\subsection*{Reliable and Secure Framework}
\begin{definition} 
        The quantity $(R_1,R_2,F,G)$ is achievable
        for the system $\mathsf{Sys}$ if there exists 
        a sequence $\{(\varphi^{(n)},$ $\psi^{(n)})\}_{n \geq 1}$
        such that 
        $\forall \delta>0$, $\exists n_0=n_0(\delta)\in\mathbb{N}_0$, 
	$\forall n\geq n_0$, 
	\begin{align*}
        &\frac {1}{n} \log |{\cal X}_i^{m_i}|
         = \frac {m_i}{n} \log |{\cal X}_i| \leq R_i, i=1,2,
        \\ 
        &p_{\rm e}(\phi^{(n)},\psi^{(n)}|P_{X_1X_2}^n) \leq 2^{-n(F-\delta)},
        \\
	&\Delta^{(n)}(\varphi^{(n)}|P_{X_1X_2}^n, P_{K_1K_2}^n)
        \leq 2^{-n(G-\delta)}.
	\end{align*}
\end{definition}

\begin{definition}{\bf (Rate Reliability and Security \\ Region)}
        Let $\mathcal{D}_{\mathsf{Sys}}({P}_{X_1X_2},$${P}_{K_1K_2})$
	denote the set of all $(R_1,R_2,F,G)$ such that
        $(R_1,R_2,F,G)$ is achievable. 
        We call $\mathcal{D}_{\mathsf{Sys}}({P}_{X_1X_2},$${P}_{K_1K_2})$
        the \emph{\bf rate reliability and security} region.
\end{definition}

\begin{definition}[Reliable and Secure Rate Region]
	We define the \emph{reliable and secure rate} region 
        $\mathcal{R}_{\mathsf{Sys}}(P_{X_1X_2},$ $P_{K_1K_2})$
        for the system $\mathsf{Sys}$ by
	\begin{align*}
	&\mathcal{R}_{\mathsf{Sys}}(P_{X_1X_2},P_{K_1K_2}):=
	\{(R_1,R_2): (R_1,R_2,F,G) 
        \\
        &\qquad \in \mathcal{D}_{\mathsf{Sys}}
        (P_{X_1X_2},P_{K_1K_2}) \mbox{ for some }F,G>0
	\}.
	\end{align*}
We call $\mathcal{R}_{\mathsf{Sys}}({P}_{X_1X_2},$${P}_{K_1K_2})$
        the \emph{\bf reliable and secure rate} region.
\end{definition}

In this paper we derive good inner bounds of 
$\mathcal{D}_{\mathsf{Sys}}($ $P_{X_1X_2},P_{K_1K_2})$ 
and $\mathcal{R}_{\mathsf{Sys}}(P_{X_1X_2},P_{K_1K_2})$.

\section{Main Results}

In this section we state our main results. To describe our results we define
several functions and sets. 
Let $\overline{X}_1$ and $\overline{X}_2$ be 
arbitrary random variables
over $\mathcal{X}_1$ and $\mathcal{X}_2$ respectively and
$P_{\overline{X}_1\overline{X}_2}$ is their joint distribution.
Let $\mathcal{P}(\mathcal{\cal X}_1\times {\cal X}_2)$ 
denote the set of all probability distributions on 
$\mathcal{X}_1\times$$\mathcal{X}_2$. Similar notations are 
adopted for other random variables. For $R\geq 0$ and 
$P_{X_1X_2} \in $ $\mathcal{P}(\mathcal{\cal X}_1\times {\cal X}_2)$, 
we define the following three functions:
\begin{align*}
	F_1(R|P_{X_1X_2}) &:{=}
	\min_{ P_{\overline{X}_1\overline{X}_2} \in 
	\mathcal{P}(\mathcal{\cal X}_1\times {\cal X}_2)}
	\{[R- H(\overline{X}_1|\overline{X}_2)]^{+}
\\
&\qquad+D(P_{\overline{X}_1\overline{X}_2}||P_{X_1X_2})\},
\\
	F_2(R|P_{X_1X_2}) &:{=}
	\min_{ P_{\overline{X}_1\overline{X}_2} \in 
	\mathcal{P}(\mathcal{\cal X}_1\times {\cal X}_2)}
	\{[R- H(\overline{X}_2|\overline{X}_1)]^{+}
\\
&\qquad+D(P_{\overline{X}_1\overline{X}_2}||P_{X_1X_2})\},
\\
	F_3(R|P_{X_1X_2}) &:{=}
	\min_{ P_{\overline{X}_1\overline{X}_2} \in 
	\mathcal{P}(\mathcal{\cal X}_1\times {\cal X}_2)}
	\{[R- H(\overline{X}_1\overline{X}_2)]^{+}
\\
&\qquad+D(P_{\overline{X}_1\overline{X}_2}||P_{X_1X_2})\},
\end{align*}
where $[a]^{+}:{=}\max \{0,a\}$. 
Furthermore, define
$$
F(R_1,R_2|P_{X_1X_2}):=\min_{i=1,2,3}F_{i}(R_i|P_{X_1X_2}),
$$
where $R_3 \defeq R_1 +R_2$. 
For random variable $Z$ with distributions $P_Z$
on finite set $\mathcal{Z}$ and any $R>0$, we define 
the following function:
\begin{align*}
G(R|P_Z):= \min_{P_{\overline{Z}}\in \mathcal{P}(\mathcal{Z})}\{
	          [H(\overline{Z})-R]^{+} + D(P_{\overline{Z}}||P_Z)\}.
\end{align*} 
For given $P_{K_1K_2}$
$\in {\cal P}({\cal X}_1\times {\cal X}_2)$, 
we define
      \begin{align*}
      	G(R_1,R_2|P_{K_1K_2})&:=\min\{
	G(R_1|P_{K_1}),G(R_2|P_{K_2}),
        \\
        &\qquad G(R_3|P_{K_1K_2})\}.
     \end{align*}
Let us define the following two regions of $(R_1,R_2)$:
\begin{align*}
 \mathcal{R}_{\mathrm{sw}}(P_{X_1X_2}):=&\{(R_1,R_2) : 
\\
& R_1 >  H(X_1|X_2), R_2 >  H(X_2|X_1), 
\\
& R_1+R_2 > H(X_1X_2)\},
\\
\mathcal{R}_{\mathrm{key}}(P_{K_1K_2}):=&\{(R_1,R_2): 
\\
& R_1 < H(K_1),R_2 < H(K_2),
\\
& R_1+R_2< H(K_1K_2)\}.
\end{align*}
Then we have the following property:
\begin{property}\label{pr:pr1}{$\quad$
\begin{itemize} 
\item[a)]
$F(R_1,R_2, P_{X_1X_2})>0$ if and only if  
$(R_1,R_2) \in $ $\mathcal{R}_{\mathrm{sw}}(P_{X_1X_2})$.
\item[b)]
$G(R_1,R_2, P_{K_1K_2})>0$ if and only if  
$(R_1,R_2) \in $ $\mathcal{R}_{\mathrm{key}}(P_{X_1X_2})$.
\end{itemize}
}
\end{property}

\newcommand{\Zasxx}{
\begin{proposition}\label{thm:error}
	For every $n$, $m_1$, $m_2$, the following holds. 
\begin{align}
	\mathbf{E}[p_{\rm e}]
	\leq 3\times 2^{\left\{-n 
         \left(\min_{i}e_i(R_i, P_{X_1X_2})-K\log(n+1)/n\right)\right
	\}},
\end{align}
where the average is taken over random choices of the linear codes encoders
$\varphi_1^{(n)}$  and $\varphi_2^{(n)}$
with uniform distributions over good sets of linear codes encoders
$\Phi(n,m_1,\mathcal{X}_1)$ and $\Phi(n,m_2,\mathcal{X}_2)$ respectively, 
$i=1,2,3$, and $K:= 2 |\mathcal {X}_1|^2| \mathcal {X}_2|^2$.
\end{proposition}
}

Our main result is as follows.
\begin{theorem}\label{Th:mainth2}{
\rm For any $R_1,R_2>0$, there exists a sequence of mappings 
$\{(\varphi^{(n)}, \psi^{(n)}) \}_{n=1}^{\infty}$
such that for any $(P_{X_1X_2},P_{K_1K_2})$ 
with $(R_1,R_2)\in $
$\mathcal{R}_{\mathrm{sw}}(P_{X_1X_2})
\cap \mathcal{R}_{\mathrm{key}}($ $P_{K_1K_2})$,
we have 
\begin{align}
& \frac {1}{n} 
\log |{\cal X}_i^{m_i}|= \frac {m_i}{n} \log |{\cal X}_i|
\leq R_i, i=1,2,
\notag\\
& 
p_{\rm e}(\phi^{(n)},\psi^{(n)}|P_{X_1X_2}^n) \leq 
2^{-n[F(R_1,R_2|P_{X_1X_2})-\delta_{1,n}]},
\label{eqn:mainThErrB}
\\
&\Delta^{(n)}(\varphi^{(n)}|P_{X_1X_2}^n,P_{K_1K_2}^n)
\leq 2^{-n[G(R_1,R_2|P_{K_1K_2})-\delta_{2,n}]}, 
\label{eqn:mainThSecB}
\end{align}
where $\delta_{i,n},i=1,2$ are defined by
\begin{align*}
\delta_{1,n}:=&
\frac{1}{n}\log\left[ 24(n+1)^{3|{\cal X}_1||{\cal X}_2|}\right],
\\
\delta_{2,n}:=& \frac{1}{n} \log \Bigl[
6(\log {\rm e})[\log(|{\cal X}_1||{\cal X}_2|)]
\\
&\quad \times n(n+1)^{3|{\cal X}_1||{\cal X}_2|}\Bigr].
\end{align*}
Note that for $i=1,2$, $\delta_{i,n} \to 0$ as $n\to \infty$. 
}
\end{theorem}

The functions $F(R_1,R_2| P_{X_1X_2})$ and $G(R_1,R_2 |P_{K_1}$ ${}_{K_2})$ 
take positive values if and only 
if $(R_1,R_2)$ $\in {\cal R}_{\rm sw}($ $P_{X_1X_2})$
$\cap{\cal R}_{\rm key}(P_{K_1K_2})$. 
Thus, by Theorem \ref{Th:mainth2}, 
under $(R_1,R_2)$ 
$\in {\cal R}_{\rm sw}($ $P_{X_1X_2})$
$\cap{\cal R}_{\rm key}(P_{K_1K_2})$, we have the followings: 
\begin{itemize}
\item On the reliability, $p_{\rm e}(\phi^{(n)},\psi^{(n)}|P_{X_1X_2}^n)$  
goes to zero exponentially as $n$ tends to infinity, and its 
exponent is lower bounded by the function $F(R_1,$ $R_2|P_{X_1X_2})$.  
\item On the security, 
$\Delta^{(n)}(\varphi^{(n)}|P_{X_1X_2}^n,P_{K_1K_2}^n)$
goes to zero exponentially as $n$ tends to infinity, and its 
exponent is lower bounded by the function $G(R_1,R_2|P_{K_1K_2})$,
\item The code that attains the exponent functions 
$F($ $R_1,R_2|P_{X_1X_2})$ and $G(R_1,R_2|P_{K_1K_2})$ 
is the universal code 
that depends only on $(R_1,R_2)$ not on the value of the distribution 
$P_{X_1X_2}$ and $P_{K_1K_2}$. 
\end{itemize}
Define
\begin{align*}
& {\cal R}_{\rm Sys}^{\rm (in)}(P_{X_1X_2},P_{K_1K_2})
:={\cal R}_{\rm sw}(P_{X_1X_2})
\cap{\cal R}_{\rm key}(P_{K_1K_2}),
\\
& {\cal D}_{\rm Sys}^{\rm (in)}(P_{X_1X_2},P_{K_1K_2})
:=\{(R_1,R_2,
\\
&\quad F(R_1,R_2|P_{X_1X_2}),G(R_1,R_2|P_{K_1K_2})):
\notag\\
& \quad (R_1,R_2) \in {\cal R}_{\rm sw}(P_{X_1X_2})
\cap{\cal R}_{\rm key}(P_{K_1K_2})\}.
\end{align*}
From Theorem \ref{Th:mainth2}, we immediately obtain 
the following corollary. 
\begin{corollary}
\begin{align*}
{\cal R}_{\rm Sys}^{\rm (in)}(P_{X_1X_2},P_{K_1K_2})
\subseteq {\cal R}_{\rm Sys}(P_{X_1X_2},P_{K_1K_2}),
\\
{\cal D}_{\rm Sys}^{\rm (in)}(P_{X_1X_2},P_{K_1K_2})
\subseteq {\cal D}_{\rm Sys}(P_{X_1X_2},P_{K_1K_2}).
\end{align*}
\end{corollary}

\begin{figure}[t]
\centering
\includegraphics[width=0.47\textwidth]{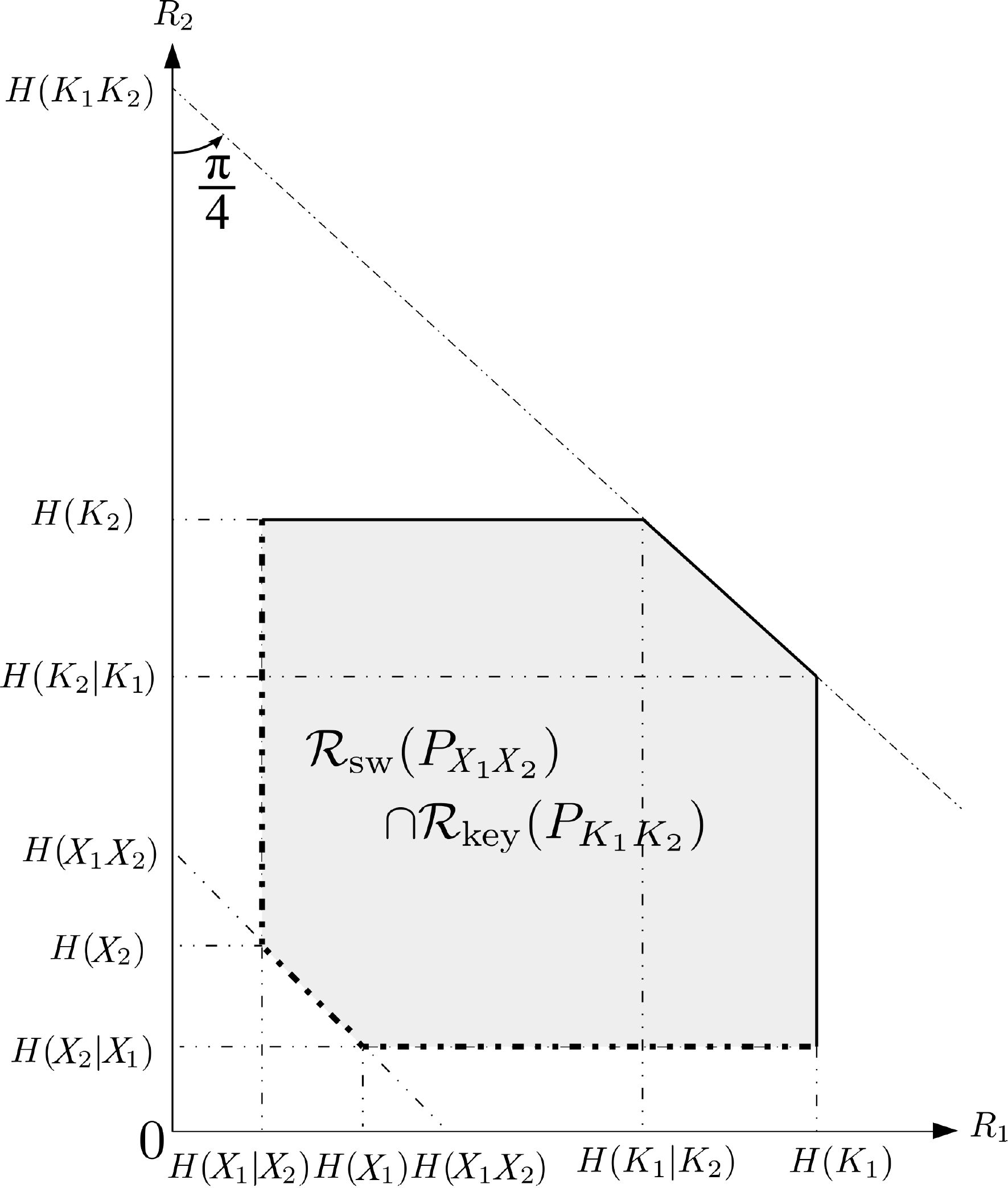}
\caption{The inner bound 
$\mathcal{R}_{\mathrm{sw}}(P_{X_1X_2})
\cap \mathcal{R}_{\mathrm{key}}(P_{K_1K_2})$
of the reliable and secure rate region 
${\cal R}_{\rm Sys}(P_{X_1X_2},$ $P_{K_1K_2})$.
}
\label{fig:admissible}
\end{figure}

A typical shape of $\mathcal{R}_{\mathrm{sw}}(P_{X_1X_2})
\cap \mathcal{R}_{\mathrm{key}}(P_{K_1K_2})$
is shown in Fig. \ref{fig:admissible}.

\newcommand{\Zss}{
}{



\section{Security Criterion Based on the Correct Probability of Decoding}


\subsection*{On Security}
An eavesdropper $\A$ who tries to estimate 
$({\rvcxone},{\rvcxtwo})\in \mathcal{X}_1^n \times \mathcal{X}_2^n$ from

$$
(\widetilde{C}_1^{m_1},\widetilde{C}_2^{m_2})
=(\varphi_1^{(n)}({\rvcxone}\Lambda_1\oplus {\rvckone}),
\varphi_2^{(n)}({\rvcxtwo}\Lambda_2\oplus {\rvcktwo}))
\in\mathcal{X}_1^{m_1}\times\mathcal{X}_2^{m_2}
$$ 
is always associated with its estimator function 
$\psi_\A$ defined by											
\begin{align}
\psi_\A:\mathcal{X}_1^{m_1}\times\mathcal{X}_2^{m_2}\rightarrow
\mathcal{X}_1^{n}\times\mathcal{X}_2^{n}.
\end{align}
For given $(P_{X_1X_2}^n, P_{K_1K_2}^n)$, 
let $p_{\rm c}(\varphi^{(n)},\psi_\A|P_{X_1X_2}^n, P_{K_1K_2}^n)$ 
denote the success probability of $\A$ correctly estimating 
$({\rvcxone},{\rvcxtwo})$ from $(Y_1^{m_1},Y_2^{m_2})$ using 
its estimation function $\psi_\A$ with respect to the pair 
of linear encoders 
$\varphi^{(n)}=(\varphi_1^{(n)}, \varphi_2^{(n)})$, 
under $(P_{X_1X_2}^n, P_{K_1K_2}^n)$.

\subsection*{Reliable and Secure Framework}
\begin{definition}
        The quantity $(R_1,R_2,F,G)$ is achievable
        for the system $\mathsf{Sys}$ if there exists 
        a sequence $\{(\varphi^{(n)},$ $\psi^{(n)})\}_{n \geq 1}$
        such that 
        $\forall \delta>0$, $\exists n_0=n_0(\delta)\in\mathbb{N}_0$, 
	$\forall n\geq n_0$, 
	\begin{align}
\frac {1}{n} \log |{\cal X}_i^{m_i}|= \frac {m_i}{n} \log |{\cal X}_i|
\leq R_i, i=1,2,\quad 				
p_{\rm e}(\phi^{(n)},\psi^{(n)}|P_{X_1X_2}^n) \leq 2^{-n(F-\delta)},
	\end{align}
	and for any eavesdropper $\A$ with $\psi_{\A}$:
	\begin{align}
	p_{\rm c}(\varphi^{(n)},\psi_{\A}|P_{X_1X_2}^n, P_{K_1K_2}^n)
        \leq 2^{-n(G-\delta)}.
	\end{align}
\end{definition}

\begin{definition}[Rate Reliablity and Security Region]
	We define the \emph{rate reliability and security} region 
        $\mathcal{\widetilde{D}}_{\mathsf{Sys}}(P_{X_1}$ ${}_{X_2},P_{K_1K_2})$
	for the system $\mathsf{Sys}$ by
	\begin{align}
	\mathcal{\widetilde{D}}_{\mathsf{Sys}}(P_{X_1X_2},P_{K_1K_2}):=
	\left\{(R_1,R_2,F,G)\:\ 
        (R_1,R_2,F,G) \mbox{ is }	
        \mbox{ achievable for }\mathsf{Sys}	
        \right\}.
	\end{align}
\end{definition}

\begin{definition}[Reliable and Secure Rate Region]
	We define the \emph{reliable and secure rate} region 
        $\mathcal{\widetilde{R}}_{\mathsf{Sys}}(P_{X_1X_2},$ $P_{K_1K_2})$
        for the system $\mathsf{Sys}$ by
	\begin{align}
	\mathcal{\widetilde{R}}_{\mathsf{Sys}}(P_{X_1X_2},P_{K_1K_2}):=
	\left\{%
		(R_1,R_2)\ \vert\ 
(R_1,R_2,F,G) \in \mathcal{\widetilde{D}}_{\mathsf{Sys}}(P_{X_1X_2},P_{K_1K_2})
                \mbox{ for some }F,G>0
	\right\}.
	\end{align}
\end{definition}

Our aim is to derive good inner bounds of 
$\mathcal{\widetilde{D}}_{\mathsf{Sys}}(P_{X_1X_2},P_{K_1K_2})$ 
and $\mathcal{\widetilde{R}}_{\mathsf{Sys}}(P_{X_1X_2},P_{K_1K_2})$.
To describe our result we define a quantity related to a correct 
probability of source estimation. 

\begin{definition}[Source Uniformity]
Let us define the following quantity:
\begin{align}
	P_{{\lrvcxone}{\lrvcxtwo}}^{*}
		&:=\max_{({\lvcxone},{\lvcxtwo})
                   \in\mathcal{X}_1^n\times\mathcal{X}_2^n} 
			P_{X_1X_2}^n({\vcxone},{\vcxtwo}).
\end{align}
Let $P_{\max}:=\max_{(x_1,x_2)\in\mathcal{X}_1\times\mathcal{X}_2} 
			P_{X_1X_2}(x_1,x_2)$.
Then, by simple computation we have
\begin{equation}
P_{{\lrvcxone}{\lrvcxtwo}}^{*}=P_{\max}^n=2^{-n \log \frac{1}{P_{\max}}}.
\end{equation}
\end{definition}

We state the following lemma which is implied directly by 
the results of Oohama \cite{DBLP:journals/corr/Oohama17b}.
\begin{lemma}\label{lem:sec_foundation} Fix positive $\nu$ arbitrary.
	In the proposed system, for any pair of encoder 
        $\varphi^{(n)}=(\varphi_1^{(n)},\varphi_2^{(n)})$,
	for any eavesdropper  $\A$ with estimator function $\psi_\A$,
	the following holds. 

	\begin{align}
	p_{\rm c}(\varphi^{(n)},\psi_\A|P^n_{X_1X_2},P^n_{K_1K_2})
	& \leq 2^{\nu}\cdot P_{{\lrvcxone}{\lrvcxtwo}}^{*}
	+\frac{1}{\nu}
         I(\widetilde{C}_1^{m_1}\widetilde{C}_2^{m_2};{\rvcxone}{\rvcxtwo}) 
	\notag\\
	&=2^{\nu}\cdot 2^{-n \log \frac{1}{P_{\max}}}
	+\frac{1}{\nu} \Delta^{(n)}(\varphi^{(n)}|P^n_{X_1X_2},P^n_{K_1K_2}).
	\label{eq:sec_foundation}
	\end{align}
\end{lemma}

Since the proof of Lemma \ref{lem:sec_foundation} is found 
in Oohama {\cite{DBLP:journals/corr/Oohama17b}, we omit the detail 
of the proof. Choosing $\nu=1$ in (\ref{eq:sec_foundation}), we have 
\begin{align}
p_{\rm c}(\varphi^{(n)},\psi_\A|P^n_{X_1X_2},P^n_{K_1K_2})
\leq 2\cdot 2^{-n \log \frac{1}{P_{\max}}}
+\Delta^{(n)}(\varphi^{(n)}|P^n_{X_1X_2},P^n_{K_1K_2}).
\label{eq:sec_foundationb}
\end{align}
From (\ref{eq:sec_foundationb}) and Theorem \ref{Th:mainth2}, we have 
the following result. 
\begin{theorem}\label{Th:mainth2b}{
\rm For any $R_1,R_2>0$, there exists a sequence of mappings 
$\{(\varphi^{(n)}, \psi^{(n)}) \}_{n=1}^{\infty}$
such that for any $(P_{X_1X_2},P_{K_1K_2})$ 
with $(R_1,R_2)\in $
$\mathcal{R}_{\mathrm{sw}}(P_{X_1X_2})
\cap \mathcal{R}_{\mathrm{key}}(P_{K_1K_2})$,
we have 
\begin{align}
& \frac {1}{n} 
\log |{\cal X}_i^{m_i}|= \frac {m_i}{n} \log |{\cal X}_i|
\leq R_i, i=1,2,
\notag\\
& 
p_{\rm e}(\phi^{(n)},\psi^{(n)}|P_{X_1X_2}^n) \leq 
2^{-n[F(R_1,R_2|P_{X_1X_2})-\delta_{1,n}]},
\label{eqn:mainThErrBb}
\end{align}
and for any eavesdropper $\A$ with $\psi_{\A}$:
	\begin{align}
		p_{\rm c}(\varphi^{(n)},\psi_{\A}|P_{X_1X_2}^n,P_{K_1K_2}^n)
        \leq 2\cdot 2^{-n \log \frac{1}{P_{\max}}}
         +
          2^{-n[G(R_1,R_2|P_{K_1K_2})-\delta_{2,n}]}, 
\label{eqn:mainThSecBb}
	\end{align}
where $\delta_{i,n},i=1,2$ are the same quantities as those 
in Theorem \ref{Th:mainth2}.
Note that for $i=1,2$, $\delta_{i,n} \to 0$ as $n\to \infty$. 
}
\end{theorem}


The functions $F(R_1,R_2| P_{X_1X_2})$ and $G(R_1,R_2|P_{K_1K_2})$ 
take positive values if and only 
if $(R_1,R_2)$ 
$\in {\cal R}_{\rm sw}($ $P_{X_1X_2})$
$\cap{\cal R}_{\rm key}(P_{K_1K_2})$. 
Thus, by Theorem \ref{Th:mainth2}, 
under $(R_1,R_2)$ 
$\in {\cal R}_{\rm sw}(P_{X_1X_2})$
$\cap{\cal R}_{\rm key}(P_{K_1K_2})$, we have the followings: 
\begin{itemize}
\item On the achievability, $p_{\rm e}(\phi^{(n)},\psi^{(n)}|P_{X_1X_2}^n)$  
goes to zero exponentially as $n$ tends to infinity, and its 
exponent is lower bounded by the function $F(R_1,R_2|P_{K_1K_2})$.  
\item On the security, for any $\psi_{\cal A}$,
$p_{\rm c}(\varphi^{(n)},\psi_{\A}|P_{X_1X_2}^n,P_{K_1K_2}^n)$
goes to zero exponentially as $n$ tends to infinity, and its 
exponent is lower bounded by the function ${G^*}(R_1,R_2|P_{K_1K_2})$,
where  
$$
G^*(R_1,R_2|P_{K_1K_2})=
\min\left\{ \log \frac{1}{P_{\max}},G(R_1,R_2|P_{K_1K_2})\right \}.
$$
\item The code that attains the exponent functions 
$F(R_1,R_2|P_{X_1X_2})$ and $G^*(R_1,R_2|P_{K_1K_2})$ 
is the universal code 
that depends only on $(R_1,R_2)$ not on the value of the distribution 
$P_{X_1X_2}$ and $P_{K_1K_2}$. 
\end{itemize}
Define
\begin{align}
\widetilde{\cal D}_{\rm Sys}^{\rm (in)}(P_{X_1X_2},P_{K_1K_2})
:=&\{(R_1,R_2,
F(R_1,R_2|P_{X_1X_2}),
G^{\ast}(R_1,R_2|P_{K_1K_2})):
\notag\\
&\:
(R_1,R_2) \in
{\cal R}_{\rm sw}(P_{X_1X_2})
\cap{\cal R}_{\rm key}(P_{K_1K_2})\}.
\end{align}
From Theorem \ref{Th:mainth2}, we immediately obtain 
the following corollary. 
\begin{corollary}
\begin{align*}
{\cal R}_{\rm Sys}^{\rm (in)}(P_{X_1X_2},P_{K_1K_2})
\subseteq \widetilde{\cal R}_{\rm Sys}(P_{X_1X_2},P_{K_1K_2}),
\\
\widetilde{\cal D}_{\rm Sys}^{\rm (in)}(P_{X_1X_2},P_{K_1K_2})
\subseteq \widetilde{\cal D}_{\rm Sys}(P_{X_1X_2},P_{K_1K_2}).
\end{align*}
\end{corollary}

\newcommand{\ZsDDD}{
\begin{figure}[t]
\centering
\includegraphics[width=0.70\textwidth]{plot-cropB.eps}
\caption{The inner bound 
$\mathcal{R}_{\mathrm{sw}}(P_{X_1X_2})
\cap \mathcal{R}_{\mathrm{key}}(P_{K_1K_2})$
of the reliable and secure rate region 
${\cal R}_{\rm Sys}^{\rm (in)}(P_{X_1X_2},P_{K_1K_2})$.
}
\label{fig:admissibleb}
\end{figure}
}


\section{Proof of the Main Result}

In this section we prove Theorems \ref{Th:mainth2} 
and \ref{Th:mainth2b}. To prove this theorem 
we use the method of types 
developed by Csisz\'ar and K\"orner \cite{csi2011information}. 
In the first subsection we prepare basic results on the types.  
Those results are basic tools for our analysis of several 
quantities related to error provability of decoding or security.
In the second subsection we evaluate upper bounds 
of $p_{\rm e}(\varphi^{(n)},\psi^{(n)}|P_{X_1X_2}^n)$ 
and $p_{\rm c}(\varphi^{(n)},\psi_{\cal A}|P_{X_1X_2}^n,P_{K_1K_2}^n)$.
We derive the upper bound $p_{\rm e}(\phi^{(n)},\psi^{(n)}|P_{X_1X_2}^n)$
which holds for any $(\phi^{(n)},\psi^{(n)})$. This result is stated 
in Lemma \ref{lem:ErBound}. 
Furthermore we derive the upper bound of 
$p_{\rm c}(\varphi^{(n)},\psi_{\mathcal{A}}|P_{X_1X_2}^n,P_{K_1K_2}^n)$
which holds for any $\varphi^{(n)}$ and 
any adversary ${\cal A}$ with $\psi_A$. This result 
is stated in Lemma \ref{lem:KeySecBound}. 
In the third subsection we develop random coding argument 
to prove an important key lemma (Lemma \ref{lem:UnivCodeBound}) 
stating an existence of good universal code 
$(\varphi^{(n)},\psi^{(n)})$. In the fourth subsection we prove 
Theorem \ref{Th:mainth2} using 
Lemma \ref{lem:ErBound},
Lemma \ref{lem:KeySecBound}, and 
Lemma \ref{lem:UnivCodeBound}. 

\subsection{Types of Sequences and Their Properties}

In this subsection we prepare basic results on the types.  
Those results are basic tools for our analysis of 
several bounds related to error provability 
of decoding or security.
\begin{definition}{\rm
For any $n$-sequence ${\vcxone}=x_{1,1}x_{1,2}\cdots $ 
$x_{1,n}\in {\calVarX}^{n}$, 
$n(x_1|{\vcxone})$ denotes the number of $t$ such that $x_{1,t}=x_1$.  
The relative frequency $\left\{n(x_1|\vcxone)/n\right\}_{x_1\in {\cal X}_1}$ 
of the components of ${\vcxone}$ is called the type of ${\vcxone}$ 
denoted by $P_{\lvcxone}$.  The set that consists of all 
the types on ${\cal X}_1$ is denoted by ${\cal P}_{n}({\cal X}_1)$. 
Let $\tX$ denote an arbitrary random variable whose distribution 
$P_{\tX}$ belongs to ${\cal P}_{n}({\cal X}_1)$. 
For $P_{\tX}\in {\cal P}_{n}({\cal X}_1)$, set 
$$
T_{\tX}^n := \left\{{\vcxone}:\, P_{{\lvcxone}}=P_{\tX}\right\}.
$$
Similarly for any two $n$-sequences 
${\vcxi}=x_{i,1}$ $x_{i,2},$ $\cdots$ $x_{i,n}\in $ 
${\cal X}_i^{n}, i=1,2$, $n(x_1,x_2|{\vcxone},{\vcxtwo})$ 
denotes the number 
of $t$ such that $(x_{1,t},$ $x_{2,t})=(x_1,$ $x_2)$.
The relative frequency 
$
\{n(x_1,x_2|{\vcxone},{\vcxtwo})/n$ $\}_{
(x_1,x_2)\in}$ ${}_{\calVarX\times\calVarY}
$ 
of the components of $({\vcxone},{\vcxtwo})$ is called the joint type 
of $({\vcxone},{\vcxtwo})$ denoted by $P_{{\vcxone},{\vcxtwo}}$. 
Furthermore, the set of all the joint type of $\calVarX\times\calVarY$ 
is denoted by ${\cal P}_{n}(\calVarX\times\calVarY)$. 
Let $(\tX,\tY)$ denote an arbitrary random pair whose distribution 
$P_{\tX\tY}$ belongs to ${\cal P}_{n}({\cal X}_1)$. 
For $P_{\tX\tY}\in {\cal P}_{n}({\calVarX\times\calVarY})$, set
$$
T^n_{\tX\tY}
:= \{({\vcxone},{\vcxtwo}):\,P_{{\lvcxone},{\lvcxtwo}}=P_{\tX\tY}
\}\,.
$$
Furthermore, for 
$P_{\tX}\in$ 
${\cal P}_{n}$ 
$({\cal X}_1)$ and 
${\vcxone} \in T_{\tX}^n$, set  
$$
T^n_{\tY|\tX}({\vcxone}) 
:= \{{\vcxtwo}:\,P_{{\lvcxone},{\lvcxtwo}}=P_{\tX\tY} \}\,.
$$
}
\end{definition}
For set of types and joint types the following lemma holds. 
For the detail of the proof see Csisz\'ar 
and K\"orner \cite{csi2011information}.
\begin{lemma}\label{lem:Lem1}{\rm 
$\quad$

\begin{itemize}
\item[a)]
   $\begin{array}[t]{l} 
   |{\cal P}_{n}({\calVarX})|\leq(n+1)^{|{\calVarX}|},  
   |{\cal P}_{n}({\calVarX}\times\calVarY)
   |\leq(n+1)^{|{\calVarX}||{\calVarY}|}.
   \end{array}$
  
\item[b)] For $P_{\tX}\in {\cal P}_{n}(\calVarX)$ and 
    $P_{\tX\tY}\in {\cal P}_{n}({\calVarX\times\calVarY})$,  
\begin{align*}
(n+1)^{-{|\calVarX|}}2^{nH(\Dist{\tX})}
&\leq |T^n_{\tX}|\leq 2^{nH(\Dist{\tX})},
\\
(n+1)^{-{|\calVarX||\calVarY|}}2^{nH({\Dist{\tX\tY}})}
&\leq |T^n_{\tX\tY}|\leq  2^{nH({\Dist{\tX\tY}})}.
\end{align*}
\item[c)]
For any ${\vcxone}\in T_{\tX}^n$, we have
$$
|T^n_{\tY|\tX}({\vcxone})|= \frac{|T^n_{\tX\tY}|}{|T^n_{\tX}|}.
$$
\item[d)] For ${\vcxone} \in T^n_{\tX}$ and $({\vcxone},{\vcxtwo})
\in T^n_{\tX\tX}$
\begin{align*}
P_{X_1}^n({\vcxone})
&=2^{-n[H(\Dist{\tX})+D(P_{\Dist{\tX}}||P_{\Dist{\VarX})}]},
P_{X_1X_2}^n({\vcxone},{\vcxtwo})
=2^{-n[H(\Dist{\tX\tY})+D(P_{\Dist{\tX\tY}}||P_\Dist{\VarX\VarX})]}.
\end{align*}
\end{itemize}
}
\end{lemma}

By Lemma \ref{lem:Lem1} parts b) and d), we immediately 
obtain the following lemma: 
\begin{lemma}\label{lem:Lem1.4}{\rm $\quad$
For $P_{\tX\tY}\in {\cal P}_{n}({\calVarX\times\calVarY})$,  
\begin{align*}
& P_{X_1X_2}^n(T_{\tX\tY}^n)
\leq 2^{-nD(P_{\Dist{\tX\tY}}||P_{\Dist{\VarX\VarY}})}.
\end{align*}
}
\end{lemma}

\subsection{Upper Bounds of 
$p_{\rm e}(\phi^{(n)},\psi^{(n)}| P_{X_1X_2}^n)$
and $p_{\rm c}(\varphi^{(n)},\psi_{\cal A}|P_{X_1X_2}^n,P_{K_1K_2}^n)$ 
}

In this subsection we evaluate upper 
bounds of $p_{\rm e}(\phi^{(n)},\psi^{(n)}| P_{X_1X_2}^n)$ and 
$p_{\rm c}(\varphi^{(n)},\psi_{\cal A}|P_{X_1X_2}^n,P_{K_1K_2}^n)$. 
For $p_{\rm e}(\phi^{(n)},\psi^{(n)}| P_{X_1X_2}^n)$, we derive 
an upper bound which can be characterized with a quantity 
depending on $(\phi^{(n)},$ $\psi^{(n)})$ and joint type 
$P_{{\lvcxone},{\lvcxtwo}}$ of sequences $(s_1^n,s_2^n) \in 
{\cal X}_1^n \times {\cal X}_2^n$.
We first evaluate $p_{\rm e}(\phi^{(n)},
\psi^{(n)}|P_{X_1X_2}^n)$.   
For $({\vcxone},{\vcxtwo}) $ $\in {\cal X}_1^{n}\times{\cal X}_2^{n}$
and $P_{\overline{X}_1\overline{X}_2}
\in{\cal P}_{n}({\cal X}_1\times{\cal X}_2)$ 
we define the following functions.
\begin{align*}
\Xi_{{\lvcxone},{\lvcxtwo}}(\phi^{(n)},\psi^{(n)})
&\defeq\left\{
\begin{array}{ccl}
1&\qquad&\mbox{if}\quad
         \psi^{(n)}\bigl(\phi^{(n)}_1({\vcxone}),
	 \phi^{(n)}_2({\vcxtwo})\bigr)
         \neq ({\vcxone},{\vcxtwo})\,,
	 \vspace{0.2cm}\\
0&\qquad&\mbox{otherwise,}
\end{array}
\right.\\
\displaystyle \Xi_{\overline{X}_1\overline{X}_2}
(\phi^{(n)},\psi^{(n)})
&\defeq \frac{1}{|T_{\overline{X}_1\overline{X}_2}^{n}|}
\sum_{({\lvcxone},{\lvcxtwo})\in T_{\overline{X}_1\overline{X}_2}^{n}}
\Xi_{{\lvcxone},{\lvcxtwo}}(\phi^{(n)},\psi^{(n)}).
\end{align*}
Then we have the following lemma. 
\begin{lemma} 
\label{lem:ErBound}
In the proposed system, for any pair of linear encoders 
$\phi^{(n)}=(\phi_1^{(n)},\phi_2^{(n)})$ and 
for any joint decoder $\psi^{(n)}$, we have
\begin{align}
& p_{\rm e}(\phi^{(n)},\psi^{(n)}|P_{X_1X_2}^n)
  \leq \sum_{P_{\overline{X}_1\overline{X}_2}
  \in {\cal P}_n({\cal X}_1\times {\cal X}_2) 
  }\Xi_{\overline{X}_1\overline{X}_2}(\phi^{(n)},\psi^{(n)})
  2^{-nD(P_{\overline{X}_1\overline{X}_2}||P_{X_1X_2})}.
\label{eqn:ErrBound}
\end{align}
\end{lemma}

\begin{IEEEproof} We have the following chain of inequalities:
\begin{align*}
&p_{\rm e}(\phi^{(n)},\psi^{(n)}|P_{X_1X_2}^n)
\MEq{a}\sum_{
P_{\overline{X}_1\overline{X}_2} 
\in {\cal P}_n({\cal X}_1\times {\cal X}_2) 
}\sum_{({\lvcxone},{\lvcxtwo})\in 
T^n_{\overline{X}_1\overline{X}_2}}
\Xi_{{\lvcxone},{\lvcxtwo}}(\phi^{(n)},\psi^{(n)})
P_{X_1X_2}^n({\vcxone},{\vcxtwo})
\\
&=\sum_{
P_{\overline{X}_1\overline{X}_2} 
\in {\cal P}_n({\cal X}_1\times {\cal X}_2)}
\frac{1}{|T^n_{\overline{X}_1\overline{X}_2}|}
\sum_{({\lvcxone},{\lvcxtwo})\in 
T^n_{\overline{X}_1\overline{X}_2}}
\Xi_{{\lvcxone},{\lvcxtwo}}(\phi^{(n)},\psi^{(n)})
|T^n_{\overline{X}_1\overline{X}_2}|
P_{X_1X_2}^n({\vcxone},{\vcxtwo})
\\
&\MEq{b}\sum_{
P_{\overline{X}_1\overline{X}_2} 
\in {\cal P}_n({\cal X}_1\times {\cal X}_2)}
\frac{1}{|T^n_{\overline{X}_1\overline{X}_2}|}
\sum_{({\lvcxone},{\lvcxtwo})\in 
T^n_{\overline{X}_1\overline{X}_2}}
\Xi_{{\lvcxone},{\lvcxtwo}}(\phi^{(n)},\psi^{(n)})
P_{X_1X_2}^n(T^n_{\overline{X}_1\overline{X}_2})
\\
&\MEq{c}\sum_{P_{\overline{X}_1\overline{X}_2} 
\in {\cal P}_n({\cal X}_1\times {\cal X}_2)}
\Xi_{\overline{X}_1\overline{X}_2}
(\phi^{(n)},\psi^{(n)})
P_{X_1X_2}^n(T^n_{\overline{X}_1\overline{X}_2})
\\
&
\MLeq{d}\sum_{P_{\overline{X}_1\overline{X}_2}
\in {\cal P}_n({\cal X}_1\times {\cal X}_2)}
\Xi_{\overline{X}_1\overline{X}_2}(\phi^{(n)},\psi^{(n)})
2^{-nD(P_{\overline{X}_1\overline{X}_2}||P_{X_1X_2})}.
\end{align*}
Step (a) follows from the definition of 
$\Xi_{{\lvcxone},{\lvcxtwo}}(\phi^{(n)},\psi^{(n)})$.
Step (b) follows from that the probabilities 
$P_{X_1X_2}^n({\vcxone},{\vcxtwo})$ for 
$({\vcxone},{\vcxtwo}) \in T^n_{\overline{X}_1\overline{X}_2}$
take an identical value.
Step (c) follows from the definition of 
$\Xi_{\overline{X}_1\overline{X}_2}(\phi^{(n)},\psi^{(n)})$.
Step (d) follows from lemma \ref{lem:Lem1.4}. 
\end{IEEEproof}
We next discuss upper bounds of 
$$
\Delta^{(n)}(\varphi^{(n)}|P_{X_1X_2}^n,P_{K_1K_2}^n)
=I(\widetilde{C}_1^{m_1},\widetilde{C}_2^{m_2};\rvcxone,\rvcxtwo).  
$$ 
On an upper bound  of 
$I(\widetilde{C}_1^{m_1},\widetilde{C}_2^{m_2};\rvcxone,\rvcxtwo)$,  
we have the following lemma.
\begin{lemma}\label{lem:MIandDiv}
\begin{align}
& I(\widetilde{C}_1^{m_1}, \widetilde{C}_2^{m_2};{\rvcxone},{\rvcxtwo})
\leq
D(P_{\widetilde{K}_1^{m_1} \widetilde{K}_2^{m_2}}||
P_{U_1^{m_1}U_2^{m_2}}), 
\label{eqn:oohama}
\end{align}
where $P_{{U_1}^{m_1}{U_2}^{m_2}}$ represents the uniform distribution
over $\mathcal{X}_1^{m_1} \times \mathcal{X}_2^{m_2}$.
\end{lemma}

\begin{IEEEproof}
We have the following chain of inequalities: 
\begin{align*}
& I(\widetilde{C}_1^{m_1}\widetilde{C}_2^{m_2};{\rvcxone},{\rvcxtwo})
=H(\widetilde{C}_1^{m_1}\widetilde{C}_2^{m_2})
-H(\widetilde{C}_1^{m_1}\widetilde{C}_2^{m_2}| {\rvcxone},{\rvcxtwo})
\\
&\MEq{a} H(\widetilde{C}_1^{m_1}\widetilde{C}_2^{m_2})
        -H(\widetilde{K}_1^{m_1}\widetilde{K}_1^{m_1}|
           {\rvcxone},{\rvcxtwo})
\MEq{b}  H(\widetilde{C}_1^{m_1}\widetilde{C}_2^{m_2})
        -H(\widetilde{K}_1^{m_1}\widetilde{K}_1^{m_1})
\\
&\leq \log (|{\cal X}_1^{m_1}||{\cal X}_2^{m_2}|)
-H(\widetilde{K}_1^{m_1}\widetilde{K}_2^{m_2})
=D(P_{\widetilde{K}_1^{m_1}\widetilde{K}_2^{m_2}}||
P_{U_1^{m_1}U_2^{m_2}}).
\end{align*}
Step (a) follows from 
$\widetilde{C}_i^{m_i}=\widetilde{K}_i^{m_i} 
    \oplus \widetilde{X}_i^{m_i}$ 
and $\widetilde{X}_i^{m_i}=\phi_i^{(n)}({\rvcxi})$
for $i=1,2$. 
Step (b) follows from 
$(\widetilde{K}_1^{m_1},\widetilde{K}_2^{m_2}) \perp ({\rvcxone},{\rvcxtwo})$. 
\end{IEEEproof}

To evaluate 
$D(P_{\widetilde{K}_1^{m_1}\widetilde{K}_2^{m_2}}||
P_{U_1^{m_1}U_2^{m_2}})$, we define the 
following quantities:
\begin{align*}
&  \Omega_{{\lvckone}, {\lvcktwo};\varphi^{(n)}}
(\widetilde{k}_1^{m_1}, \widetilde{k}_2^{k_2})
:=\left\{
\begin{array}[c]{l}
    1, \mbox{ if } (\varphi_1^{(n)}({\vckone}),
                    \varphi_2^{(n)}({\vcktwo})
                     =(\widetilde{k}_1^{m_1},
                       \widetilde{k}_2^{m_2}),\\
    0, \mbox{ otherwise }
\end{array}\right.
\\
&\Omega_{\overline{K}_1\overline{K}_2;\varphi^{(n)}}
(\widetilde{k}_1^{m_1},\widetilde{k}_2^{m_2})
:=\frac{1}{|T^n_{\overline{K}_1\overline{K}_2}|}
\sum_{({\lvckone},{\lvcktwo})\in T_{\overline{K}_1\overline{K}_2}^n} 
\Omega_{{\lvckone},{\lvcktwo};\varphi^{(n)}}
(\widetilde{k}_1^{m_1},\widetilde{k}_2^{m_2}).
\end{align*}
From the above definition, we can regard
$
\left\{
\Omega_{\overline{K}_1\overline{K}_2;\varphi^{(n)}}
(\widetilde{k}_1^{m_1},\widetilde{k}_2^{m_2})
\right\}_{
(\widetilde{k}_1^{m_1},\widetilde{k}_2^{m_2})
\in {\cal X}_1^{m_1}\times {\cal X}_2^{m_2}}
$
as a probability distribution on 
${\cal X}_1^{m_1} \times {\cal X}_2^{m_2}$. 
We denote this probability distribution 
by $\Omega_{\overline{K}_1\overline{K}_2;\varphi^{(n)}}$. 
By the definition of 
$\Omega_{\overline{K}_1\overline{K}_2;\varphi^{(n)}}$, 
for $P_{\overline{K}_1\overline{K}_2}$ 
$\in {\cal P}_n({\cal X}_1\times {\cal X}_2)$, we have the following:
\begin{align}
& P_{\widetilde{K}_1^{m_1} \widetilde{K}_2^{m_2}}
(\widetilde{k}_1^{m_1},\widetilde{k}_2^{m_2})=
\sum_{P_{\overline{K}_1\overline{K}_2} 
\in {\cal P}_n({\cal X}_1\times {\cal X}_2) }
\Omega_{\overline{K}_1\overline{K}_2;\varphi^{(n)}}
(\widetilde{k}_1^{m_1},\widetilde{k}_2^{m_2})
P^n_{\overline{K}_1\overline{K}_2}
(T_{\overline{K}_1\overline{K}_2}^n)
\label{eqn:sssR}\\
& \mbox{ for } (\widetilde{k}_1^{m_1},\widetilde{k}_2^{m_2}) 
\in {\cal X}_1^{m_1} \times {\cal X}_2^{m_2}.
\nonumber
\end{align}
Furthermore, we define 
$$
\Delta_{\overline{K}_1\overline{K}_2}(\varphi^{(n)})
:= \sum_{(\widetilde{k}_1^{m_1},\widetilde{k}_2^{m_2})
\in {\cal X}_1^{m_1} \times {\cal X}_2^{m_2}}
|{\cal X}_1|^{m_1} |{\cal X}_2|^{m_2}
\left(
\Omega_{\overline{K}_1\overline{K}_2;\varphi^{(n)}}
( \widetilde{k}_1^{m_1}, \widetilde{k}_2^{m_2})
-\frac{1}{|{\cal X}_1|^{m_1}|{\cal X}_2|^{m_2}}
\right)^2.
$$
Then we have the following lemma.
\begin{lemma}\label{lem:LemDiv}
\begin{align}
&D(P_{\widetilde{K}_1^{m_1}
   \widetilde{K}_2^{m_2}}||P_{U_1^{m_1}U_2^{m_2}})
=D(P_{\varphi_1^{(n)}(K_1^{n})\varphi_2^{(n)}(K_2^{n})}
||P_{U_1^{m_1}U_2^{m_2}})
\notag\\
&
\leq (\log{\rm e})[\log(|{\cal X}_1||{\cal X}_2|)]n \cdot
       \sum_{
       P_{\overline{K}_1\overline{K}_2} 
       \in {\cal P}_n({\cal X}_1\times{\cal X}_2)}
\Delta^{\ast}_{\overline{K}_1\overline{K}_2}(\varphi^{(n)})
2^{-nD(P_{\overline{K}_1\overline{K}_2}||P_{K_1K_2})},
\label{eqn:SzzzP}
\end{align}
where
$
\Delta^{\ast}_{\overline{K}_1\overline{K}_2}(\varphi^{(n)})
:=\min\{1,\Delta_{\overline{K}_1\overline{K}_2}(\varphi^{(n)})\}.
$
\end{lemma}

\begin{IEEEproof} By (\ref{eqn:sssR}) and the convexity 
of divergence we have
\begin{align}
&D(P_{\widetilde{K_1}^{m_1}\widetilde{K_2}^{m_2}}
  ||P_{U_1^{m_1}U_2^{m_2}})
\leq \sum_{
(\widetilde{k}_1^{m_1},\widetilde{k}_2^{m_2})
\in {\cal X}_1^{m_1}\times {\cal X}_2^{m_2}}
       \sum_{P_{\overline{K}_1\overline{K}_2 } 
         \in {\cal P}_n({\cal X}\times{\cal Y})}
             P^n_{\overline{K}_1\overline{K}_2}
(T_{\overline{K}_1\overline{K}_2}^n)
\notag\\
&\quad \times
\Omega_{\overline{K}_1\overline{K}_2;\varphi^{(n)}}
 (\widetilde{k}_1^{m_1},\widetilde{k}_2^{m_2})
\log\left(|{\cal X}_1|^{m_1}|{\cal X}_2|^{m_2}
\Omega_{\overline{K}_1\overline{K}_2;\varphi^{(n)}}
 (\widetilde{k}_1^{m_1},\widetilde{k}_2^{m_2})
\right)
\nonumber\\
&=\sum_{
       P_{\overline{K}_1\overline{K}_2} 
       \in {\cal P}_n({\cal X}_1\times{\cal X}_2)}
       P^n_{K_1K_2}
       (T_{\overline{K}_1\overline{K}_2}^n)
       D(\Omega_{\overline{K}_1\overline{K}_2;\varphi^{(n)}}
         ||P_{U_1^{m_1}U_2^{m_2}})
\nonumber\\
&\MLeq{a} \sum_{
       P_{\overline{K}_1\overline{K}_2} 
       \in {\cal P}_n({\cal X}_1\times{\cal X}_2)}
       D(\Omega_{\overline{K}_1\overline{K}_2;\varphi^{(n)}}
         ||P_{U_1^{m_1}U_2^{m_2}})
        2^{-nD(P_{\overline{K}_1\overline{K}_2}||P_{K_1K_2})}.
\label{eqn:PrEq3}
\end{align}
Step (a) follows from Lemma \ref{lem:Lem1.4}. Hence, it suffices 
to derive an upper bound 
of $D(\Omega_{\overline{K}_1\overline{K}_2;\varphi^{(n)}}
    ||$ $P_{U_1^{m_1}U_2^{m_2}})$.
Since 
$$
D(\Omega_{\overline{K}_1\overline{K}_2;\varphi^{(n)}}
||P_{U_1^{m_1}U_2^{m_2}})
=m_1 \log |{\cal X}_1|+ m_2\log |{\cal X}_2|
-H(\Omega_{\overline{K}_1\overline{K}_2;\varphi^{(n)}})
\leq [\log(|{\cal X}_1||{\cal X}_2|)]n,
$$
$D(\Omega_{\overline{K}_1\overline{K}_2;\varphi^{(n)}}
||P_{U_1^{m_1}U_2^{m_2}})$ 
has the obvious upper bound $[\log(|{\cal X}_1||{\cal X}_2|)]n$.
Note that this quantity is larger than $n$. We next derive another
upper bound of 
$D(\Omega_{\overline{K}_1\overline{K}_2;\varphi^{(n)}}
||P_{U_1^{m_1}U_2^{m_2}})$.
Using that the inequality
$$ 
u\log(u/v)\leq \left(\log{\rm e}\right)\left\{u-v+(u-v)^2/v \right\}
$$ 
holds for any positive number $u,v$, we obtain 
\begin{align}
&D(\Omega_{\overline{K}_1\overline{K}_2;\varphi^{(n)}})
    ||P_{U_1^{m_1}U_2^{m_2}})
\leq \left(\log{\rm e}\right)
\sum_{(\widetilde{k}_1^{m_1},\widetilde{k}_2^{m_2})
\in {\cal X}_1^{m_1}\times {\cal X}_2^{m_2}
}|{\cal X}_1|^{m_1}|{\cal X}_2|^{m_2}
\nonumber\\
&\qquad \times \left(
\Omega_{\overline{K}_1\overline{K}_2;\varphi^{(n)}}
(\widetilde{k}_1^{m_1},\widetilde{k}_2^{m_2})
-\frac{1}{|{\cal X}_1|^{m_1}|{\cal X}_2|^{m_2}}
\right)^2
=\left(\log{\rm e}\right)
\Delta_{\overline{K}_1\overline{K}_2}(\varphi^{(n)}).
\label{eqn:PrEq4}
\end{align}
From (\ref{eqn:PrEq4}) and the upper bound 
$[\log(|{\cal X}_1||{\cal X}_2|)]n$ of 
$D(\Omega_{\overline{K}_1\overline{K}_2;\varphi^{(n)}} 
||P_{U_1^{m_1}U_2^{m_2}})$ larger than $n$,
we have
\begin{align}
D(\Omega_{\overline{K}_1\overline{K}_2;\varphi^{(n)}}
||P_{U_1^{m_1}U_2^{m_2}}) 
&\leq \left(\log{\rm e}\right)[\log(|{\cal X}_1||{\cal X}_2|)]
n \min\{1,\Delta_{\overline{K}_1\overline{K}_2}(\varphi^{(n)})\}
\notag\\
&=\left(\log{\rm e}\right)[\log(|{\cal X}_1||{\cal X}_2|)]n
\Delta^{\ast}_{\overline{K}_1\overline{K}_2}(\varphi^{(n)}). 
\label{eqn:PrEq5}
\end{align}
Combining (\ref{eqn:PrEq3}) and (\ref{eqn:PrEq5}), we have 
the bound (\ref{eqn:SzzzP}) of Lemma \ref{lem:LemDiv}. 
\end{IEEEproof}

Combining Lemma \ref{lem:MIandDiv} and Lemma \ref{lem:LemDiv},
we have the following lemma. 
\begin{lemma}\label{lem:KeySecBound}
	In the proposed system, for any pair of encoder 
        $\varphi^{(n)}=(\varphi_1^{(n)},\varphi_2^{(n)})$,
	for any eavesdropper  $\A$ with estimator function $\psi_\A$,
	we have
	\begin{align}
       &\Delta^{(n)}(\varphi^{(n)}|P^n_{X_1X_2},P^n_{K_1K_2})
        \notag\\
       &\leq  
	\left(\log{\rm e}\right)
        [\log(|{\cal X}_1||{\cal X}_2|)]n\cdot
       \sum_{
       P_{\overline{K}_1\overline{K}_2} 
       \in {\cal P}_n({\cal X}_1\times{\cal X}_2)}
       \Delta^{\ast}_{\overline{K}_1\overline{K}_2}(\varphi^{(n)})
       2^{-nD(P_{\overline{K}_1\overline{K}_2}||P_{K_1K_2})}.
\label{eqn:SecBound}
\end{align}
\end{lemma}

The bound (\ref{eqn:ErrBound}) in Lemma \ref{lem:ErBound} 
implies that upper bounds of  
$\Xi_{\overline{X}_1\overline{X}_2}(\phi^{(n)},\psi^{(n)})$ 
for $P_{\overline{X}_1\overline{X}_2}$ 
lead to derivations of good error bounds on 
$p_{\rm e}(\phi^{(n)},\psi^{(n)}|P_{X_1X_2}^n)$.
Furthermore, the bound (\ref{eqn:SecBound}) in 
Lemma \ref{lem:KeySecBound} 
implies that good upper bounds of  
$\Delta_{\overline{K}_1\overline{K}_2}(\varphi^{(n)})$ 
for $P_{\overline{K}_1\overline{K}_2}$ 
lead to derivations of good secure upper bounds 
on $p_{\rm c}(\varphi^{(n)},\psi_{\cal A}|P_{X_1X_2}^n,P_{K_1K_2}^n$ $)$.
In the next subsection we discuss an existence of 
universal code 
$$
(\varphi^{(n)},\psi^{(n)})
=(\varphi_1^{(n)},\varphi_2^{(n)},\psi^{(n)})
=(\phi_1^{(n)} \oplus a_1^{m_1}, \phi_2^{(n)}\oplus a_2^{m_2},\psi^{(n)})
$$ 
such that the quantities 
$\Xi_{\overline{X}_1\overline{X}_2}
(\phi^{(n)},\psi^{(n)})$ 
for $P_{\overline{X}_1\overline{X}_2}\in 
{\cal P}_n({\cal X}_1\times {\cal X}_2)$ and 
$\Delta^{\ast}_{\overline{K}_1\overline{K}_2}(\varphi^{(n)})$ 
for $P_{\overline{K}_1\overline{K}_2}$ 
$\in {\cal P}_n({\cal X}_1$ $\times {\cal X}_2)$ 
attain the bound of Theorem \ref{Th:mainth2}.

\subsection{Random Coding Arguments}

We construct a pair of affine encoders 
$\varphi^{(n)}=(\varphi_1^{(n)},\varphi_2^{(n)})$ 
using the random coding method. For the joint decoder 
$\psi^{(n)}$, we propose the minimum entropy decoder 
used in Csisz\'{a}r \cite{Csiszr1982LinearCF} and Oohama and Han 
\cite{DBLP:journals/tit/OohamaH94}.
 
\noindent
\underline{\it Random Construction of Affine Encoders:} \  
We first choose $m_i,i=1,2$ such that 
$$
m_i:=\left\lfloor\frac{nR_i}{\log |{\cal X}_i|}\right\rfloor,
$$
where $\lfloor a \rfloor$ stands for the integer part of $a$. 
It is obvious that for $i=1,2$, we have  
$$
R_i-\frac{1}{n} \leq \frac{m_i}{n}\log |{\cal X}_i|\leq R_i. 
$$
By the definition (\ref{eq:homomorphica}) of 
$\phi_i^{(n)},i=1,2$, we have that for each $i=1,2$ and for  
${\vcxi} \in {\cal X}_i^n$,
\begin{align*}
& \phi_i^{(n)}({\vcxi} {})={\vcxi} {} A_i,
\end{align*}
where $A_i$ is a matrix with $n$ rows and $m_i$ columns.
By the definition (\ref{eq:homomorphic}) of 
$\varphi_i^{(n)},i=1,2$, we have that for each $i=1,2$ and 
for ${\vcki} \in {\cal X}_i^n$,
\begin{align*}
& \varphi_i^{(n)}({\vcki})={\vcki} A_i+b_i^{m_i},
\end{align*}
where $b_i^{m_i}$ is a vector with $m_i$ columns.
For each $i=1,2$, entries of $A_i$ and $b_i^{m_i}$ 
are from the field of ${\cal X}_i$. Those entries 
are selected at random, 
independently of each other and 
with uniform distribution.
Randomly constructed  
linear encoders $\phi_i^{(n)},i=1,2$ and affine encoders 
and $\varphi_i^{(n)},i=1,2$ have three properties shown 
in the following lemma.
\newcommand{\LemmaAffine}{
\begin{lemma}[Properties of Linear/Affine Encoders]
\label{lem:good_set}
$\quad$
\begin{itemize}
\item[a)] For each $i=1,2$, and for 
$\vcxi, \vcvi \in {\cal X}_i^n$.
we set ${\vcyi} \defeq {\vcxi}{},
{\vcwi} \defeq {\vcvi}{}$.
Then we have the followings. For each $i=1,2$, and 
for any ${\vcxi}, {\vcvi} \in {\cal X}_i^n$ with 
         ${\vcxi} \neq {\vcvi}$, we have
	\begin{align} 
       	 \Pr[\phi_i^{(n)}({\vcxi}{})=
         \phi_i^{(n)}({\vcvi}{})]=
         \Pr[({\vcxi} \ominus {\vcvi}) A_i =0^{m_i}]=|\mathcal{X}_i|^{-m_i}.
	\end{align}

\item[b)] For each $i=1,2$, for any ${\vcsi} \in {\cal X}_i^n$, 
         and for any $\widetilde{s}_i^{m_i}\in {\cal X}_i^{m_i}$, we have
	\begin{align} 
	  \Pr[\varphi_i^{(n)}({\vcsi})=\widetilde{s}_i^{m_i}]=
          \Pr[ {\vcsi}A_i \oplus b_i^{m_i}=\widetilde{s}_i^{m_i}]
         =|\mathcal{X}_i|^{-m_i}.
	\end{align}
\item[c)] 
For each $i=1,2$, 
for any ${\vcsi}, {\vcti} \in {\cal X}_i^n$ 
with ${\vcsi} \neq {\vcti}$, 
and for any $\widetilde{s}_i^{m_i}, \widetilde{t}_i^{m_i} 
            \in {\cal X}_i^{m_i}$, we have
	\begin{align} 
	&\Pr[\varphi_i^{(n)}({\vcsi}) =\widetilde{s}_i^{m_i}
            \land  \varphi_i^{(n)}({\vcti})=\widetilde{t}_i^{m_i}]
        \notag\\
        &= \Pr[{\vcsi} A_i \oplus b_i^{m_i}=\widetilde{s}_i^{m_i}
               \land 
               {\vcti} A_i \oplus b_i^{m_i}=\widetilde{t}_i^{m_i}]  
         = |\mathcal{X}_i|^{-2m_i}.
	\end{align}
\end{itemize}
\end{lemma}
}
\newcommand{\LemmaAffineB}{
\begin{lemma}[Properties of Linear/Affine Encoders]
\label{lem:good_set}
$\quad$
\begin{itemize}
\item[a)] 
For each $i=1,2$, and 
for any ${\vcxi}, {\vcvi} \in {\cal X}_i^n$ with 
         ${\vcxi} \neq {\vcvi}$, we have
	\begin{align} 
       	 \Pr[\phi_i^{(n)}({\vcxi}{})=
         \phi_i^{(n)}({\vcvi}{})]=
         \Pr[({\vcyi} \ominus {\vcwi}) A_i =0^{m_i}]=|\mathcal{X}_i|^{-m_i}.
	\end{align}

\item[b)] For each $i=1,2$, for any ${\vcsi} \in {\cal X}_i^n$, 
         and for any $\widetilde{s}_i^{m_i}\in {\cal X}_i^{m_i}$, we have
	\begin{align} 
	  \Pr[\varphi_i^{(n)}({\vcsi})=\widetilde{s}_i^{m_i}]=
          \Pr[ {\vcsi}A_i \oplus b_i^{m_i}=\widetilde{s}_i^{m_i}]
         =|\mathcal{X}_i|^{-m_i}.
	\end{align}
\item[c)] 
For each $i=1,2$, 
for any ${\vcsi}, {\vcti} \in {\cal X}_i^n$ 
with ${\vcsi} \neq {\vcti}$, 
and for any $\widetilde{s}_i^{m_i}
\in {\cal X}_i^{m_i}$, we have
	\begin{align} 
	&\Pr[\varphi_i^{(n)}({\vcsi})=
            \varphi_i^{(n)}({\vcti})=\widetilde{s}_i^{m_i}]
        = \Pr[{\vcsi} A_i \oplus b_i^{m_i}=
               {\vcti} A_i \oplus b_i^{m_i}=\widetilde{s}_i^{m_i}]  
         = |\mathcal{X}_i|^{-2m_i}.
	\end{align}
\end{itemize}
\end{lemma}
}
\LemmaAffineB

Proof of this lemma is given in Appendix \ref{apd:ProofLemAA}. 
\newcommand{\ProofLemAA}{
\subsection{Proof of Lemma \ref{lem:good_set}}
\label{apd:ProofLemAA}

In this appendix we prove Lemma \ref{lem:good_set}. The suffix $i$ 
in ${\cal X}_i$ used to distinguish ${\cal X}_1$ and ${\cal X}_2$ 
in Lemma \ref{lem:good_set} is not essential for the proof. 
In the following argument we omit this suffix. Let ${\cal X}$ 
be a finite field and let $\Lambda$ be an $n\times n$ invertible 
matrix, whose entries are from ${\cal X}$.  
Let $\phi: {\cal X}^n \to {\cal X}^m$ be a linear 
map with $\phi({\vcx}\Lambda)={\vcx} \Lambda A$ 
for ${\vcx} \in {\cal X}^n$. 
Here $A$ is a matrix with $n$ rows and 
$m$ colomus. 
Let $\varphi: {\cal X}^n \to {\cal X}^m$ be an affine map with 
$\varphi(\vcs)= \vcs A \oplus b^m$ for $\vcs \in {\cal X}^n$. 
Here $b^m$ is a vectorx with $m$ colomus. 
Entries of $A$ and $b^m$ are from the field 
of ${\cal X}$. Those entries are selected at random, 
independently of each other and 
with uniform distribution. In this appendix we prove 
the following lemma. 
\begin{lemma}\label{lem:good_setb}
$\quad$
\begin{itemize}
\item[a)]  
For any ${\vcx}, {\vcv} \in {\cal X}^n$ with $ {\vcx} \neq {\vcv}$,
we have
	\begin{align} 
		\Pr[\phi({\vcx})=\phi({\vcv})]=
                \Pr[({\vcx} \ominus {\vcv})A=0^m]=|\mathcal{X}|^{-m}.
	\end{align}
\item[b)] For any ${\vcs} \in {\cal X}^n$, 
         and for any $\widetilde{s}^{m}\in {\cal X}^m$, we have
	\begin{align} 
	 \Pr[\varphi({\vcs})=\widetilde{s}^{m}]=
         \Pr[{\vcs} A \oplus b^{m}=\widetilde{s}^{m}]
         =|\mathcal{X}|^{-m}.
	\end{align}
\item[c)] 
For any ${\vcs}, {\vct} \in {\cal X}^n$ with ${\vcs} \neq {\vct}$, 
and for any $\widetilde{s}^{m}
            \in {\cal X}^{m}$, we have
	\begin{align} 
	&\Pr[\varphi({\vcs})=\varphi({\vct})=\widetilde{s}^{m}]
        =\Pr[{\vcs} A \oplus b^{m}=
               {\vct} A \oplus b^{m}=\widetilde{s}^{m}]
         =|\mathcal{X}|^{-2m}.
	\end{align}
\newcommand{\ZZZ}{
For any ${\vcs}, {\vct} \in {\cal X}^n$ with ${\vcs} \neq {\vct}$, 
and for any $\widetilde{s}^{m}, \widetilde{t}^{m} 
            \in {\cal X}^{m}$, we have
	\begin{align} 
	&\Pr[\varphi({\vcs})=\widetilde{s}^{m}
            \land \varphi({\vct})=\widetilde{t}^m]
        \notag\\
        &= \Pr[{\vcs} A \oplus b^{m}=\widetilde{s}^{m}
               \land 
               {\vct} A \oplus b^{m}=\widetilde{t}^{m}]  
         =|\mathcal{X}|^{-2m}.
	\end{align}
}
\end{itemize}	
\end{lemma}

\begin{IEEEproof}
Let $a_l^m$ be the $l$-th low vector of the matrix $A$.
For each $l=1,2,\cdots,n$, let $A_l^m \in{\cal X}^m$ 
be a random vector which represents 
the randomness of the choice of 
$a_l^m \in{\cal X}^m$. 
Let $B^m \in{\cal X}^m$ be a random vector which represent 
the randomness of the choice of $b^m \in{\cal X}^m$.
We first prove the part a).  
Since $\Lambda$ is invertible, we have 
$$
{\vcx} \neq {\vcv} \Leftrightarrow x_i \neq v_i
\mbox{ for some } i \in \{1,2,\cdots,n\}. 
$$
Without loss of generality we may assume 
$x_1 \neq v_1$. 
Under this assumption we have the following:
\begin{align}
({\vcx}\ominus \vcv)A=0^m \Leftrightarrow 
\sum_{l=1}^n(x_l \ominus v_l)a_l^m=0^m 
\Leftrightarrow a_1^m=\sum_{l=2}^n 
\frac{w_l \ominus x_l}{x_1 \ominus w_1}a_l^m.
\label{eqn:Awff}
\end{align}
Computing $\Pr[\phi({\vcx})=\phi({\vcv})]$,
we have the following chain of equalities:
\begin{align*}
 &\Pr[\phi({\vcx})=\phi({\vcv})]
=\Pr[(\vcx \ominus \vcv )A=0^m]
\MEq{a}\Pr\left[ a_1^m=\sum_{l=2}^n 
\frac{v_l\ominus x_l}{x_1 \ominus v_1}a_l^m \right]
\\
&\MEq{b}
\sum_{ \scs \left\{a_l^m \right\}_{l=2}^n 
        \atop{\scs 
        \in {\cal X}^{(n-1)m}
        }
    }   
\prod_{l=2}^n P_{A_l^m}(a_l^m)      
P_{A_1^m}\left(\sum_{l=2}^n 
\frac{w_l\ominus y_l}{y_1 \ominus w_1}a_l^m \right)
=|\mathcal{X}|^{-m}
\sum_{ \scs \left\{a_l^m \right\}_{l=2}^n 
        \atop{
        \scs \in {\cal X}^{(n-1)m}
       }
    }   
\prod_{l=2}^n P_{A_l^m}(a_l^m)      
=|\mathcal{X}|^{-m}.
\end{align*} 
Step (a) follows from (\ref{eqn:Awff}). 
Step (b) follows from that $n$ random vecotors 
$A_l^m, l=1,2,\cdots,n$ are independent. 
We next prove the part b). We have 
the following:
\begin{align}
{\vcs} A \oplus b^{m}=\widetilde{s}^{m}
\Leftrightarrow 
b^m=\widetilde{s}^{m} \ominus 
\left\{ \sum_{l=1}^n s_la_l^m \right\}. 
\label{eqn:Awffb}
\end{align}
Computing $\Pr[{\vcs} A \oplus b^{m}=\widetilde{s}^{m}]$,
we have the following chain of equalities:
\begin{align*}
&\Pr[{\vcs} A \oplus b^{m}=\widetilde{s}^{m}]
\MEq{a}\Pr\left[b^m=\widetilde{s}^{m} \ominus 
 \left\{ \sum_{l=1}^n s_la_l^m\right\} 
\right]
\MEq{b} 
\sum_{ \scs \left\{a_l^m \right\}_{l=1}^n 
       \atop{\scs 
        \in {\cal X}^{nm}
       }
    }   
\prod_{l=1}^n P_{A_l^m}(a_l^m)      
P_{B^m}\left(
\widetilde{s}^{m} \ominus 
 \left\{ \sum_{l=1}^n s_la_l^m\right\}
\right)
\\
&=|\mathcal{X}|^{-m}
  \sum_{ \scs \left\{a_l^m \right\}_{l=1}^n 
        \atop{\scs 
        \in {\cal X}^{nm}
        }
    }   
\prod_{l=1}^n P_{A_l^m}(a_l^m)=|\mathcal{X}|^{-m}. 
\end{align*} 
Step (a) follows from (\ref{eqn:Awffb}). 
Step (b) follows from that 
$n$ random vecotors $A_l^m, l=1,2, \cdots,n$ 
and $B^m$ are independent. 
We finally prove the part c). We first observe that
$
{\vcs} \neq {\vct} \Leftrightarrow 
$
is equivalent to 
$
s_i \neq t_i 
\mbox{ for some } i \in \{1,2,\cdots,n\}. 
$
Without loss of generality, we may assume that
$s_1 \neq t_1$. Under this assumption we have 
the following:
\newcommand{\Argument}{
\begin{align}
&\left.
\ba{l}
{\vcs} A \oplus b^{m}=\widetilde{s}^{m},
\\
{\vct} A \oplus b^{m}=\widetilde{t}^{m}
\ea
\right\}
\Leftrightarrow 
\left.
\ba{l}
({\vcs}\ominus {\vct})A=\widetilde{s}^m\ominus\widetilde{t}^m, 
\\
b^m=\ds \widetilde{s}^{m} \ominus 
\left\{\sum_{l=1}^n s_la_l^m \right\} 
\ea
\right\}
\notag\\
&\Leftrightarrow 
\left.
\ba{l}
\ds a_1^m=\sum_{l=2}^n 
   \frac{t_l \ominus s_l}{s_1 \ominus t_1} a_l^m
   \oplus 
\frac{\widetilde{s}^m \ominus \widetilde{t}^m}{s_1\ominus t_1},
\\
\ds b^m= \widetilde{s}^{m} \ominus 
\left\{ \sum_{l=1}^n s_l a_l^m \right\} 
\ea
\right\}
\Leftrightarrow 
\left.
\ba{l}
\ds a_1^m=\sum_{l=2}^n 
   \frac{t_l \ominus s_l}{s_1 \ominus t_1} a_l^m
   \oplus 
\frac{\widetilde{s}^m \ominus \widetilde{t}^m}{s_1\ominus t_1},
\\
\ds b^m=
\sum_{l=2}^n
\frac{t_1s_l\ominus s_1t_l}
{s_1\ominus t_1}a_l^m
\ominus \frac{t_1\widetilde{s}^m  \ominus s_1 \widetilde{t}^m }
{s_1\ominus t_1} 
\ea
\right\}.
\label{eqn:Awffc}
\end{align}
Then we have the following chain of equalities:
\begin{align*}
&\Pr[ {\vcs} A \oplus b^{m}=\widetilde{s}^{m}
\land {\vct} A \oplus b^{m}=\widetilde{t}^{m}]
\\
&\MEq{a}
\Pr\left[
a_1^m=\sum_{l=2}^n\frac{t_l \ominus s_l}{s_1 \ominus t_1} a_l^m
\oplus \frac{\widetilde{s}^m \ominus \widetilde{t}^m}{s_1\ominus t_1}
\land
b^m=
\sum_{l=2}^n
\frac{t_1s_l \ominus s_1t_l}
{s_1\ominus t_1}a_l^m
\ominus \frac{t_1\widetilde{s}^m  \ominus s_1 \widetilde{t}^m }
{s_1\ominus t_1} 
\right]
\\
&\MEq{b}
\sum_{ \scs \left\{ a_l^m \right \}_{l=2}^n 
      \atop{\scs
        \in {\cal X}^{(n-1)m}
      }
    }   
\prod_{l=2}^n P_{A_l^m}(a_l^m)      
P_{A_1^m}\left(\sum_{l=2}^n\frac{t_l \ominus s_l}{s_1 \ominus t_1} a_l^m
\oplus \frac{\widetilde{s}^m \ominus \widetilde{t}^m}{s_1\ominus t_1}
\right)
\\
&\qquad \times P_{B^m}
\left(
\sum_{l=2}^n
\frac{t_1s_l\ominus s_1t_l}
{s_1\ominus t_1}a_l^m
\ominus \frac{t_1\widetilde{s}^m  \ominus s_1 \widetilde{t}^m }
{s_1\ominus t_1} 
\right)
=|\mathcal{X}|^{-2m}
 \sum_{ \scs \left\{a_l^m \right\}_{l=2}^n 
        \atop{\scs 
        \in {\cal X}^{(n-1)m}
        }
    }   
\prod_{l=2}^n P_{A_l^m}(a_l^m)      
=|\mathcal{X}|^{-2m}.
\end{align*}
Step (a) follows from (\ref{eqn:Awffc}). 
Step (b) follows from the independent 
property on $A_l^m, l=1,2,\cdots,n$ and $B^m.$
}
\newcommand{\ArgumentB}{
\begin{align}
&
\ba{l}
{\vcs} A \oplus b^{m}=
{\vct} A \oplus b^{m}=\widetilde{s}^{m}
\ea
\Leftrightarrow 
\ba{l}
({\vcs}\ominus {\vct})A=0,
b^m=\ds \widetilde{s}^{m} \ominus 
\left\{\sum_{l=1}^n s_la_l^m \right\} 
\ea
\notag\\
&\Leftrightarrow 
\ba{l}
\ds a_1^m=\sum_{l=2}^n 
   \frac{t_l \ominus s_l}{s_1 \ominus t_1} a_l^m,
\ds b^m= \widetilde{s}^{m} \ominus 
\left\{ \sum_{l=1}^n s_l a_l^m \right\} 
\ea
\notag\\
&
\Leftrightarrow 
\ba{l}
\ds a_1^m=\sum_{l=2}^n 
   \frac{t_l \ominus s_l}{s_1 \ominus t_1} a_l^m,
\ds b^m= \widetilde{s}^m \oplus 
\sum_{l=2}^n
\frac{t_1s_l\ominus s_1t_l}{s_1\ominus t_1}a_l^m.
\ea
\label{eqn:Awffc}
\end{align}
Computing 
$\Pr[{\vcs} A \oplus b^{m}={\vct} A \oplus b^{m}=\widetilde{s}^{m}]$,
we have the following chain of equalities:
\begin{align*}
&\Pr[{\vcs} A \oplus b^{m}={\vct} A \oplus b^{m}=\widetilde{s}^{m}]
\\
&\MEq{a}
\Pr\left[
a_1^m=\sum_{l=2}^n\frac{t_l \ominus s_l}{s_1 \ominus t_1} a_l^m
\land
b^m=\widetilde{s}^m \oplus 
\sum_{l=2}^n
\frac{t_1s_l \ominus s_1t_l}
{s_1\ominus t_1}a_l^m
\right]
\\
&\MEq{b}
\sum_{ \scs \left\{ a_l^m \right \}_{l=2}^n 
      \atop{\scs
        \in {\cal X}^{(n-1)m}
      }
    }   
\left[\prod_{l=2}^n P_{A_l^m}(a_l^m)\right]      
P_{A_1^m}\left(\sum_{l=2}^n\frac{t_l \ominus s_l}{s_1 \ominus t_1} a_l^m
\right)
P_{B^m}
\left(
\widetilde{s}^m \oplus
\sum_{l=2}^n
\frac{t_1s_l\ominus s_1t_l}
{s_1\ominus t_1}a_l^m
\right)
\\
&=|\mathcal{X}|^{-2m}
 \sum_{ \scs \left\{a_l^m \right\}_{l=2}^n 
        \atop{\scs 
        \in {\cal X}^{(n-1)m}
        }
    }   
\prod_{l=2}^n P_{A_l^m}(a_l^m)      
=|\mathcal{X}|^{-2m}.
\end{align*}
Step (a) follows from (\ref{eqn:Awffc}). 
Step (b) follows from the independent 
property on $A_l^m, l=1,2,\cdots,n$ and $B^m.$
}
\ArgumentB
\end{IEEEproof}
}
We next define the joint decoder function 
$\psi^{(n)}: 
{\cal X}_1^{m_1} \times {\cal X}_2^{m_2}
\to {\cal X}_1^{n} \times {\cal X}_2^{n}.$
To this end we define the following quantities.   
\begin{definition}{\rm
 For $({\vcxone},{\vcxtwo})
 \in{\cal X}_1^{n}\times {\cal X}_2^{n}$,
we denote the conditional entropy and entropy calculated
from the joint type $P_{{\lvcxone},{\lvcxtwo}}$ by 
$H({\vcxone}|{\vcxtwo})$ and $H({\vcxone}{\vcxtwo})$, 
respectively. In other words, for a joint type  
$P_{\overline{X}_1\overline{X}_2}
\in {\cal P}_n({\cal X}_1\times {\cal X}_2)$ 
such that $P_{\overline{X}_1\overline{X}_2}
=P_{{\lvcxone},{\lvcxtwo}}$, 
we define 
$H({\vcxone}|{\vcxtwo})=H(\overline{X}_1|\overline{X}_2)$ and
$H({\vcxone} {\vcxtwo})=H(\overline{X}_1\overline{X}_2)$.}
\end{definition}

\noindent
\underline{\it Minimum Entropy Decoder:} \ For 
$\phi_i^{(n)}(x_i^n {})=\widetilde{y}_i^{m_i},i=1,2$, 
we define the joint decoder function 
$
\psi^{(n)}: {\cal X}_1^{m_1}\times {\cal X}_2^{m_2}
\to {\cal X}_1^{n} \times {\cal X}_2^{n}
$ as follows:
$$
\psi^{(n)}(\widetilde{x}_1^{m_1},\widetilde{x}_2^{m_2})
:=\left\{\begin{array}{cl}
({\hvcxone}, {\hvcxtwo})
   &\mbox{if } \phi_1^{(n)}({\hvcxone}{})=\widetilde{x}_1^{m_1},
               \phi_2^{(n)}({\hvcxtwo}{})=\widetilde{x}_2^{m_2},\\
   &\mbox{and }H({\hvcxone} {\hvcxtwo})
    <H( {\cvcxone} {\cvcxtwo})\\
   &\mbox{for all }({\cvcxone},{\cvcxtwo})\mbox{ such that }\\
   & \:\phi_1^{(n)}({\cvcxone}{})=\widetilde{x}_1^{m_1},
       \phi_2^{(n)}({\cvcxtwo}{})=\widetilde{x}_2^{m_2},\\
   & \mbox{and } 
    \:({\cvcxone},{\cvcxtwo})
    \neq ({\hvcxone},{\hvcxtwo}),
\vspace{0.2cm}\\
\mbox{arbitrary}
    & \mbox{if there is no such }({\hvcxone},{\hvcxtwo})
     \in{\cal X}_1^{n}\times{\cal X}_2^{n}.
\end{array}
\right.$$

\noindent
\underline{\it Error Probability Bound:} \ In the following 
arguments we let expectations and variances based on 
the randomness of the encoder functions be denoted by 
${\bf E}$ $[$ $\cdot]$ and ${\bf Var}[\cdot]$, 
respectively. Define 
$$
\Psi_{\overline{X}_1\overline{X}_2}(R_1,R_2):=
    2^{-n[R_1-H(\overline{X}_1|\overline{X}_2)]^{+}}
  + 2^{-n[R_2-H(\overline{X}_2|\overline{X}_1)]^{+}}
  + 2^{-n[R_1+R_2-H(\overline{X}_1\overline{X}_2)]^{+}}.
$$
Then we have the following lemma.
\begin{lemma}\label{lem:LemA}
For any $n$ and for any
$P_{\overline{X}_1\overline{X}_2}
\in{\cal P}_{n}({\cal X}_1 \times {\cal X}_2)$,
$$
{\bf E}\left[
\Xi_{\overline{X}_1\overline{X}_2}(\phi^{(n)},\psi^{(n)})
\right]
\leq
4(n+1)^{|{\cal X}_1||{\cal X}_2|}
\Psi_{\overline{X}_1\overline{X}_2}(R_1,R_2).
$$
\end{lemma}

Proof of this lemma is given in Appendix \ref{apd:ProofLemA}.
\newcommand{\ProofLemA}{
\subsection{Proof of Lemma \ref{lem:LemA}}
\label{apd:ProofLemA}

For simplicity of notation, we write 
$M_i=|{\cal X}_i|^{m_i},i=1,2.$ 
We also use those notations in the arguments of 
other appendixes. 

\begin{IEEEproof}[Proof of Lemma \ref{lem:LemA}] 
For ${{\vcxone}}\in {\cal X}_1^{n},{{\vcxtwo}}\in{\cal X}_2^{n}$ we set 
\begin{align}
B({\vcxone}{{\vcxtwo}})&=
\Bigl\{\,({\cvcxone},{\cvcxtwo}):\,
H({\cvcxone} {\cvcxtwo})
\leq H({{\vcxone}}{{\vcxtwo}})\,,\:
P_{{\clvcxone}}=P_{{\lvcxone}},
P_{{\clvcxtwo}}=P_{{\lvcxtwo}}\,
\Bigr\},
\nonumber\\
B({{\vcxone}}|{{\vcxtwo}})&=
\Bigl\{\,{\cvcxone}:\,
H({\cvcxone}|{\vcxtwo})
\leq H({{\vcxone}}|{\vcxtwo}),\:
P_{{\clvcxone}}=P_{{\lvcxone}}\,\Bigr\},
\nonumber\\
B({{\vcxtwo}}|{{\vcxone}})&=
\Bigl\{\,{\cvcxtwo}:\, 
H({\cvcxtwo}|{\vcxone})
\leq H({\vcxtwo}|{{\vcxone}}),\:
P_{{\clvcxtwo}}=P_{{\lvcxtwo}}\,\Bigr\}.
\nonumber
\end{align}
Using parts a) and b) of Lemma \ref{lem:Lem1}, we have 
following inequalities:
\begin{align}
|B({\vcxone}{\vcxtwo})|
&\leq (n+1)^{|{\cal X}_1||{\cal X}_2|}2^{nH({\lvcxone} {\lvcxtwo})},
\label{eqn:b1}\\
|B({\vcxone}|{\vcxtwo})|
&\leq (n+1)^{|{\cal X}_1||{\cal X}_2|}2^{nH({\lvcxone}|{\lvcxtwo})},
\label{eqn:b2}\\
|B({\vcxtwo}|{\vcxone})|
&\leq (n+1)^{|{\cal X}_1||{\cal X}_2|}2^{H({\lvcxtwo}| {\lvcxone})}.
\label{eqn:b3}
\end{align}
On an upper bound of ${\bf E}[\Xi_{{{\lvcxone}},
{{\lvcxtwo}}}(\phi^{(n)},\psi^{(n)})]$,
we have the following chain of inequalities:
\begin{align*}
&  {\bf E}[\Xi_{{{\lvcxone}},{{\lvcxtwo}}}(\phi^{(n)},\psi^{(n)})]
\nonumber\\
&\leq\sum_{\scs
      {\clvcxone} \in B({{\lvcxone}}|{{\lvcxtwo}}),
       \atop{\scs 
             {\clvcxone} \neq {\lvcxone} 
            }
     }
\Pr\bigl\{\phi_1^{(n)}({\cvcxone}{})
          =\phi_1^{(n)}({\vcxone}{})\bigr\}
    +\sum_{\scs 
          {\clvcxtwo} \in B({\lvcxtwo}|{\lvcxone}),
     \atop{\scs {\clvcxtwo} \neq {\lvcxtwo} 
     }
}
\Pr\bigl\{\phi_2^{(n)}({\cvcxtwo}{})=
          \phi_2^{(n)}({{\vcxtwo}}{})\bigr\}
\nonumber\\
&\quad+\sum_{\scs ({\clvcxone},{\clvcxtwo})\in 
            B({\lvcxone} {\lvcxtwo} ),
           \atop{\scs
             {\clvcxone} \neq {\lvcxone},
             {\clvcxtwo} \neq {\lvcxtwo}
          }
     }
\Pr\bigl\{
\phi_1^{(n)}({\cvcxone}{})=\phi_1^{(n)}({\vcxone}{}),
\phi_2^{(n)}({\cvcxtwo}{})=\phi_2^{(n)}({\vcxtwo}{})
\bigr\}
\nonumber\\
&\MLeq{a}\sum_{ {\clvcxone} \in 
      B({\lvcxone}|{\lvcxtwo})}\frac{1}{M_1}
     +\sum_{{\clvcxtwo} \in 
       B({\lvcxtwo}|{\lvcxone})}\frac{1}{M_2}
     +\sum_{({\clvcxone},{\clvcxtwo})\in 
     B({\lvcxone} {\lvcxtwo})}\frac{1}{M_1M_2}
\nonumber\\
&=  \frac{|B({\vcxone}|{\vcxtwo})|}{M_1}
     +\frac{|B({\vcxtwo}|{\vcxone})|}{M_2}
     +\frac{|B({\vcxone}{\vcxtwo})|}{M_1M_2}
\\
& \MLeq{b}4(n+1)^{|{\cal X}_1||{\cal X}_2|}
\left\{
 2^{-n[R_1-H({{\lvcxone}}|{{\lvcxtwo}})]}
+2^{-n[R_2-H({{\lvcxtwo}}|{{\lvcxone}})]}
+2^{-n[R_1+R_2-H({{\lvcxone}}{{\lvcxtwo}})]}\right\}.
\end{align*}
Step (a) follows from Lemma \ref{lem:good_set} part a) 
and independent random constructions of linear 
encoders $\phi_1^{(n)}$ and $\phi_2^{(n)}$.
Step (b) follows from (\ref{eqn:b1}), (\ref{eqn:b2}), 
(\ref{eqn:b3}), and $M_i \geq 2^{nR_i-1}, i=1,2$. 
On the other hand we have the obvious bound 
${\bf E}[\Xi_{{{\lvcxone}},{{\lvcxtwo}}}(\phi^{(n)},\psi^{(n)})]\leq 1$.
Hence we have 
\begin{align*}
&{\bf E}[\Xi_{{{\lvcxone}},{{\lvcxtwo}}}(\phi^{(n)},\psi^{(n)})]
\\
& \leq 4(n+1)^{|{\cal X}_1||{\cal X}_2|}
\left\{
2^{-n[R_1-H({{\lvcxone}}|{{\lvcxtwo}})]^{+}}
+2^{-n[R_2-H({{\lvcxtwo}}|{{\lvcxone}})]^{+}}
+2^{-n[R_1+R_2-H({{\lvcxone}}{{\lvcxtwo}})]^{+}}
\right\}.
\end{align*}
Hence we have
\begin{align*}
&{\bf E}[\Xi_{\overline{X}_1\overline{X}_2}(\phi^{(n)},\psi^{(n)})]
={\bf E}
\left[
\frac{1}{|T^n_{\overline{X}_1\overline{X}_2}|}
\sum_{({\lvcxone},{\lvcxtwo})
\in T^n_{\overline{X}_1\overline{X}_2}}
\Xi_{{{\lvcxone}},{{\lvcxtwo}}}(\phi^{(n)},\psi^{(n)})
\right]
\\
&= \frac{1}{|T^n_{\overline{X}_1\overline{X}_2}|}
  \sum_{({\lvcxone},{\lvcxtwo}) \in T^n_{\overline{X}_1\overline{X}_2}}
 {\bf E}[\Xi_{{\lvcxone},{\lvcxtwo}}(\phi^{(n)},\psi^{(n)})]
\\
&\leq 4(n+1)^{|{\cal X}_1||{\cal X}_2|}
\left\{
 2^{-n[R_1-H(\overline{X}_1|\overline{X}_2)]^{+}}
+2^{-n[R_2-H(\overline{X}_2|\overline{X}_1)]^{+}}
+2^{-n[R_1+R_2-H(\overline{X}_1\overline{X}_2)]^{+}}
\right\},
\end{align*}
completing the proof.
\end{IEEEproof}
}

\noindent
\underline{\it Estimation of Approximation Error:} \ 
Define 
$$
\Theta_{\overline{K}_1\overline{K}_2}(R_1,R_2)
:=2^{-n[H(\overline{K}_1)-R_1]}
+ 2^{-n[H(\overline{K}_2)-R_2]}
+ 2^{-n[H(\overline{K}_1\overline{K}_2)-(R_1+R_2)]}.
$$
Then we have the following lemma.
\begin{lemma}\label{lem:LemB} \ \ For any 
$\widetilde{k}_i^{m_i}\in {\cal X}_i^{m_i},$ $i=1,2$
\begin{align}
\E[\Omega_{\overline{K}_1\overline{K}_2;\varphi^{(n)}}
(\widetilde{k}_1^{m_1},\widetilde{k}_2^{m_2})]
&=\frac{1}{|{\cal X}_1|^{m_1}|{\cal X}_2|^{m_2}},
\label{eqn:Lem2a}
\\
\Var[\Omega_{\overline{K}_1
\overline{K}_2;\varphi^{(n)}}(\widetilde{k}_1^{m_1},\widetilde{k}_2^{m_2})]
&\leq 
  \frac{(n+1)^{|{\cal X}_1||{\cal X}_2|}}
{|{\cal X}_1|^{2m_1}|{\cal X}_2|^{2m_2}}
 \cdot \Theta_{\overline{K}_1\overline{K}_2}(R_1,R_2).
\label{eqn:Lem2b}
\end{align}
\end{lemma}

Proof of this lemma is given in Appendix \ref{apd:ProofLemB}. 
\newcommand{\ProofLemB}{
\subsection{Proof of Lemma \ref{lem:LemB}}
\label{apd:ProofLemB}

\begin{IEEEproof}[Proof of Lemma \ref{lem:LemB}]
We first compute the expectation of
$\Omega_{\overline{K}_1\overline{K}_2;\varphi^{(n)}}
(\widetilde{k}_1^{m_1},\widetilde{k}_2^{m_2})$. 
We obtain the following:  
\begin{align}
& {\bf E}\left[\Omega_{\overline{K}_1\overline{K}_2;\varphi^{(n)}}
(\widetilde{k}_1^{m_1},\widetilde{k}_2^{m_2})\right]
 ={\bf E}\left[
\frac{1}{|T^n_{\overline{K}_1\overline{K}_2}|}
\sum_{({\lvckone},{\lvcktwo}) 
\in T_{\overline{K}_1\overline{K}_2}^n} 
\Omega_{{\lvckone},{\lvcktwo};\varphi^{(n)}}
(\widetilde{k}_1^{m_1},\widetilde{k}_2^{m_2})
\right]
\notag\\
&= \frac{1}{|T^n_{\overline{K}_1\overline{K}_2}|}
\sum_{({\lvckone},{\lvcktwo}) 
\in T_{\overline{K}_1\overline{K}_2}^n} 
{\bf E}\left[\Omega_{{\lvckone},{\lvcktwo};\varphi^{(n)}}
(\widetilde{k}_1^{m_1},\widetilde{k}_2^{m_2})\right]
\MEq{a}
\frac{1}{|T^n_{\overline{K}_1\overline{K}_2}|}
\sum_{({\lvckone},{\lvcktwo}) 
\in T_{\overline{K}_1\overline{K}_2}^n} 
\frac{1}{M_1M_2}=\frac{1}{M_1M_2}.
\notag
\end{align}
Step (a) follows from Lemma \ref{lem:good_set} part b)
and independent random constructions of affine 
encoders $\varphi_1^{(n)}$ and
$\varphi_2^{(n)}$.
Thus (\ref{eqn:Lem2a}) is proved. Next, we prove (\ref{eqn:Lem2b}).
We have the following chain of equalities:   
\begin{align}
& |T_{\overline{K}_1\overline{K}_2}^n|^2{\bf E}\left[
          \left(\Omega_{\overline{K}_1\overline{K}_2}
    (\widetilde{k}_1^{m_1},\widetilde{k}_2^{m_2})\right)^2 
       \right]
\nonumber\\
&=\sum_{({\lvckone},{\lvcktwo})
     \in T_{\overline{K}_1\overline{K}_2}^n}
    \sum_{({\hlvckone},{\hlvcktwo})\in T_{\overline{K}_1\overline{K}_2}^n}
        {\bf E}\left[
        \Omega_{\lvckone,\lvcktwo;\varphi^{(n)}}
        (\widetilde{k}_1^{m_1},\widetilde{k}_2^{m_2}) 
        \Omega_{{\hlvckone},{\hlvcktwo};\varphi^{(n)}}
        (\widetilde{k}_1^{m_1},\widetilde{k}_2^{m_2})
        \right]
\nonumber\\
&\MEq{a}\sum_{({\lvckone},{\lvcktwo})\in T_{\overline{K}_1\overline{K}_2}^n}
       \frac{1}{M_1M_2}
+\sum_{\scs {\lvckone} \in T_{\overline{K}_1}^n,
          \atop{\scs
             {\lvcktwo} \ne {\hlvcktwo}
         \in T_{\overline{K}_2|\overline{K}_1}^n(\lvckone)
            } 
          }
        \frac{1}{M_1M_2^2}
+\sum_{\scs {\lvcktwo} \in T_{\overline{K}_2}^n,
          \atop{\scs
             {\lvckone} \ne {\hlvckone} 
             \in T_{\overline{K}_1|\overline{K}_2}^n(\lvcktwo)
            } 
          }
        \frac{1}{M_1^2M_2}
+\sum_{\scs({\lvckone},{\lvcktwo}) \in T_{\overline{K}_1\overline{K}_2}^n,
           \atop{\scs ({\hlvckone},{\hlvcktwo})
           \in T_{\overline{K}_1\overline{K}_2}^n,
                 \atop{\scs
                  {\lvckone} \ne {\hlvckone},
                  {\lvcktwo} \ne {\hlvcktwo}
                }
           }
        }
        \frac{1}{M_1^2M_2^2}
\nonumber\\
&\leq 
\frac{ |T_{\overline{K}_1\overline{K}_2}^n|}{M_1M_2}
+\sum_{\scs {\lvckone} \in T_{\overline{K}_1}^n}
        \frac{|T_{\overline{K}_2|\overline{K}_1}^n(\vckone)|^2}{M_1M_2^2}
+\sum_{\scs {\lvcktwo} \in T_{\overline{K}_2}^n}
        \frac{|T_{\overline{K}_1|\overline{K}_2}^n(\vcktwo)|^2}{M_1^2M_2}
+       \frac{|T_{\overline{K}_1\overline{K}_2}^n|^2}{M_1^2M_2^2}
\nonumber\\
&\MEq{b} \frac{|T_{\overline{K}_1\overline{K}_2}^n|}{M_1M_2}
        +\frac{|T_{\overline{K}_1}^n|
               |T_{\overline{K}_1\overline{K}_2}^n|^2
               }{|T_{\overline{K}_1}^n|^2 M_1M_2^2} 
        +\frac{|T_{\overline{K}_2}^n|
               |T_{\overline{K}_1\overline{K}_2}^n|^2
               }{|T_{\overline{K}_2}^n|^2 M_1^2M_2} 
        +\frac{|T_{\overline{K}_1\overline{K}_2}^n|^2}{M_1^2M_2^2}
\nonumber\\
&=  \frac{|T_{\overline{K}_1\overline{K}_2}^n|^2}{M_1^2M_2^2}
         \left\{
         \frac{M_1M_2}{|T_{\overline{K}_1\overline{K}_2}^n|}
        +\frac{M_1} {|T_{\overline{K}_1}^n|}
        +\frac{M_2} {|T_{\overline{K}_2}^n|}
        +1 \right\}.
\label{eqn:prlm1}
\end{align}
Step (a) follows from Lemma \ref{lem:good_set} part c) 
and independent random constructions of 
affine encoders $\varphi_1^{(n)}$ and
$\varphi_2^{(n)}$. Step (b) follows from Lemma \ref{lem:Lem1} 
part c). Hence we have 
\begin{align*}
& {\bf Var}\left[
      \left(
      \Omega_{\overline{K}_1\overline{K}_2;\varphi^{(n)}}
      (\widetilde{k}_1^{m_1},\widetilde{k}_2^{m_2})
      \right)^2\right]
={\bf E}\left[
   \left(\Omega_{\overline{K}_1\overline{K}_2;\varphi^{(n)}}
      (\widetilde{k}_1^{m_1},\widetilde{k}_2^{m_2})\right)^2
   \right]
-\left({\bf E}\left[
   \Omega_{\overline{K}_1\overline{K}_2;\varphi^{(n)}}
       (\widetilde{k}_1^{m_1},\widetilde{k}_2^{m_2})
   \right]\right)^2
\nonumber\\
&\leq  \frac{1}{M_1^2M_2^2}\left\{ 
        \frac{M_1M_2}{|T_{\overline{K}_1\overline{K}_2}^n|}
        + \frac{M_1}{|T_{\overline{K}_1}^n|}
        + \frac{M_2}{|T_{\overline{K}_2}^n|}
        \right\}
\MLeq{a} \frac{1}{M_1^2M_2^2}
       \left\{
           \frac{2^{n(R_1+R_2)}}
           {|T_{\overline{K}_1\overline{K}_2}^n|}
         + \frac{2^{nR_1}}{|T_{\overline{K}_1}^n|}
         + \frac{2^{nR_2}}{|T_{\overline{K}_2}^n|}
       \right\}
\\
&\MLeq{b} 
\frac{(n+1)^{|{\cal X}_1||{\cal X}_2|}}{M_1^2M_2^2}
\left\{ 
  2^{-n[H(\overline{K}_1\overline{K}_2)-(R_1+R_2)]}
+ 2^{-n[H(\overline{K}_1)-R_1]}
+ 2^{-n[H(\overline{K}_2)-R_2]}
\right\}.
\end{align*}
Step (a) follows from $M_i\leq 2^{nR_i}, i=1,2$. 
Step (b) follows from Lemma \ref{lem:Lem1} part b). 
\end{IEEEproof}
}
We have the following corollary from Lemma \ref{lem:LemB}. 
\begin{corollary}\label{cor:Szz}
For any $P_{\overline{K}_1\overline{K}_2}
\in {\cal P}_n({\cal X}_1\times{\cal X}_2)$, 
we have
$$
{\bf E}\left[\Delta_{\overline{K}_1\overline{K}_2}
(\varphi^{(n)})
\right]
\leq (n+1)^{|{\cal X}_1||{\cal X}_2|}
\Theta_{\overline{K}_1\overline{K}_2}(R_1,R_2).
$$
\end{corollary}

\noindent
\underline{\it Existence of 
Good Universal Code $(\varphi^{(n)},\psi^{(n)})$:} \ From 
Lemma \ref{lem:LemA} and Corollary \ref{cor:Szz}, we have 
the following lemma stating an existence 
of good universal code $(\varphi^{(n)},\psi^{(n)})$. 

\begin{lemma}\label{lem:UnivCodeBound} There exists at 
least one deterministic code  
$(\varphi^{(n)},\psi^{(n)})$ satisfying 
$(m_i/n)\log |{\cal X}_i|\leq R_i,i=1,2$, such that 
for any 
$P_{\overline{X}_1\overline{X}_2}$, 
$P_{\overline{K}_1\overline{K}_2}$ 
$\in {\cal P}_n({\cal X}_1 \times {\cal X}_2)$,
\begin{align*}
\Xi_{\overline{X}_1\overline{X}_2}(\phi^{(n)},\psi^{(n)})
\leq 
8(n+1)^{2|{\cal X}_1||{\cal X}_2|}
\Psi_{\overline{X}_1\overline{X}_2}(R_1,R_2),
\\
\Delta_{\overline{K}_1\overline{K}_2}
(\varphi^{(n)})
\leq 2(n+1)^{2|{\cal X}_1||{\cal X}_2|}
\Theta_{\overline{K}_1\overline{K}_2}(R_1,R_2).
\end{align*}
\end{lemma}

\begin{IEEEproof}
We have the following chain of inequalities:
\begin{align}
&{\bf E}
\left[
\sum_{
P_{\overline{X}_1\overline{X}_2} 
\in {\cal P}_n({\cal X}_1\times {\cal X}_2)
}
\left(4(n+1)^{|{\cal X}_1||{\cal X}_2|}
\Psi_{\overline{X}_1\overline{X}_2}(R_1,R_2) \right)^{-1}
\Xi_{\overline{X}_1\overline{X}_2}(\phi^{(n)},\psi^{(n)})
\right.
\notag\\
&\qquad \left. 
+\sum_{
 P_{\overline{K}_1 \overline{K}_2}
\in {\cal P}_n({\cal X}_1 \times {\cal X}_2)
}
\left(
(n+1)^{|{\cal X}_1||{\cal X}_2|}
\Theta_{\overline{K}_1\overline{K}_2}(R_1,R_2)
\right)^{-1}
\Delta_{ \overline{K}_1 \overline{K}_2 }
(\varphi^{(n)})
\right]
\notag\\
&=
\sum_{
P_{\overline{X}_1\overline{X}_2} 
\in {\cal P}_n({\cal X}_1\times {\cal X}_2)
}
\left(4(n+1)^{|{\cal X}_1||{\cal X}_2|}
\Psi_{\overline{X}_1\overline{X}_2}(R_1,R_2) \right)^{-1}
{\bf E}\left[
\Xi_{\overline{X}_1\overline{X}_2}(\phi^{(n)},\psi^{(n)})
\right]
\notag\\
&\qquad 
+\sum_{
 P_{\overline{K}_1 \overline{K}_2}
\in {\cal P}_n({\cal X}_1 \times {\cal X}_2)
}
\left((n+1)^{|{\cal X}_1||{\cal X}_2|}
\Theta_{\overline{K}_1\overline{K}_2}(R_1,R_2)
\right)^{-1}
{\bf E}\left[\Delta_{ \overline{K}_1 \overline{K}_2 }
(\varphi^{(n)})
\right]
\notag\\
&\MLeq{a} 
\sum_{
 P_{\overline{K}_1 \overline{K}_2}
\in {\cal P}_n({\cal X}_1 \times {\cal X}_2)
}1
+
\sum_{
P_{\overline{X}_1\overline{X}_2} 
\in {\cal P}_n({\cal X}_1\times {\cal X}_2)
}1
\MLeq{b} 2|{\cal P}_n({\cal X}_1 \times {\cal X}_2)|
 \leq 2(n+1)^{|{\cal X}_1||{\cal X}_2|}.
\notag
\end{align}
Step (a) follows from Lemma \ref{lem:LemA} and 
Corollary \ref{cor:Szz}. Step (b) follows 
from Lemma \ref{lem:Lem1} part a).
Hence there exists at least one deterministic code
$(\varphi^{(n)},\psi^{(n)})$ such that
\begin{align}
&\sum_{
P_{\overline{X}_1\overline{X}_2} 
\in {\cal P}_n({\cal X}_1\times {\cal X}_2)
}
\left(4(n+1)^{|{\cal X}_1||{\cal X}_2|}
\Psi_{\overline{X}_1\overline{X}_2}(R_1,R_2) \right)^{-1}
\Xi_{\overline{X}_1\overline{X}_2}(\phi^{(n)},\psi^{(n)})
\notag\\
+&\sum_{
 P_{\overline{K}_1 \overline{K}_2}
\in {\cal P}_n({\cal X}_1 \times {\cal X}_2)
}
\left((n+1)^{|{\cal X}_1||{\cal X}_2|}
\Theta_{\overline{K}_1\overline{K}_2}(R_1,R_2)
\right)^{-1}
\Delta_{\overline{K}_1 \overline{K}_2}(\varphi^{(n)})
\leq 2(n+1)^{|{\cal X}_1||{\cal X}_2|},
\notag
\end{align}
from which we have that 
\begin{align}
& \left(4(n+1)^{|{\cal X}_1||{\cal X}_2|}
 \Psi_{\overline{X}_1\overline{X}_2}(R_1,R_2) \right)^{-1}
 \Xi_{\overline{X}_1\overline{X}_2}(\phi^{(n)},\psi^{(n)})
 \leq 2(n+1)^{|{\cal X}_1||{\cal X}_2|},
\notag\\
&\left(
(n+1)^{|{\cal X}_1||{\cal X}_2|}
\Theta_{\overline{K}_1\overline{K}_2}(R_1,R_2)
\right)^{-1}
\Delta_{ \overline{K}_1 \overline{K}_2 }
(\varphi^{(n)})
\leq 2(n+1)^{|{\cal X}_1||{\cal X}_2|}
\notag
\end{align}
for any 
$P_{\overline{X}_1\overline{X}_2}$, 
$P_{\overline{K}_1\overline{K}_2}$ 
$\in {\cal P}_n({\cal X}_1 \times {\cal X}_2)$.
Thus we have 
\begin{align}
& \Xi_{\overline{X}_1\overline{X}_2}(\phi^{(n)},\psi^{(n)})
 \leq 8(n+1)^{2|{\cal X}_1||{\cal X}_2|}
 \Psi_{\overline{X}_1\overline{X}_2}(R_1,R_2), 
\notag\\
& \Delta_{ \overline{K}_1 \overline{K}_2 }
(\varphi^{(n)})
\leq 2(n+1)^{2|{\cal X}_1||{\cal X}_2|}
\Theta_{\overline{K}_1\overline{K}_2}(R_1,R_2)
\notag
\end{align}
for any 
$P_{\overline{X}_1\overline{X}_2}$, 
$P_{\overline{K}_1\overline{K}_2}$ 
$\in {\cal P}_n({\cal X}_1 \times {\cal X}_2)$.
\end{IEEEproof}

\noindent
\subsection{Proof of Theorem \ref{Th:mainth2}} 

In this subsection we prove Theorem \ref{Th:mainth2} using 
Lemma \ref{lem:ErBound}, Lemma \ref{lem:KeySecBound}, 
and Lemma \ref{lem:UnivCodeBound}. 

\begin{IEEEproof}[Proof of Theorem \ref{Th:mainth2}]
By Lemma \ref{lem:UnivCodeBound}, there exists 
$(\varphi^{(n)},\psi^{(n)})$ satisfying 
$(m_i/n)\log |{\cal X}_i|\leq R_i,i=1,2$, such that 
for any 
$P_{\overline{X}_1\overline{X}_2}$, 
$P_{\overline{K}_1\overline{K}_2}$ 
$\in {\cal P}_n({\cal X}_1 \times {\cal X}_2)$,
\begin{align}
\Xi_{\overline{X}_1\overline{X}_2}(\phi^{(n)},\psi^{(n)})
\leq 
8(n+1)^{2|{\cal X}_1||{\cal X}_2|}
\Psi_{\overline{X}_1\overline{X}_2}(R_1,R_2),
\label{eqn:aaSSS}
\\
\Delta_{\overline{K}_1\overline{K}_2}
(\varphi^{(n)})
\leq 2(n+1)^{|{\cal X}_1||{\cal X}_2|}
\Theta_{\overline{K}_1\overline{K}_2}(R_1,R_2).
\label{eqn:abSSS}
\end{align}
We first prove (\ref{eqn:mainThErrB}) in Theorem \ref{Th:mainth2}.
On an upper bound of $p_{\rm e}(\phi^{(n)},\psi^{(n)}|P_{X_1X_2}^n)$,  
we have the following chain of inequalities:  
\begin{align*}
& p_{\rm e}(\phi^{(n)},\psi^{(n)}|P_{X_1X_2}^n)  
\MLeq{a} 8(n+1)^{2|{\cal X}_1||{\cal X}_2|}
   \sum_{P_{\overline{X}_1\overline{X}_2}
   \in {\cal P}_n({\cal X}_1\times {\cal X}_2) }
   \Psi_{\overline{X}_1\overline{X}_2}(R_1,R_2)
   2^{-nD(P_{\overline{X}_1\overline{X}_2}||P_{X_1X_2})}
\\
&\MLeq{b}24(n+1)^{2|{\cal X}_1||{\cal X}_2|}
|{\cal P}_n({\cal X}_1\times {\cal X}_2)|
2^{-n[\min_{i=1,2,3} F_i(R_i|P_{X_1X_2})]} 
\\
&\MLeq{c} 24(n+1)^{3|{\cal X}_1||{\cal X}_2|}
  2^{-nF(R_1,R_2|P_{X_1X_2})}
 =2^{-n[F(R_1,R_2|P_{X_1X_2})-\delta_{1,n}]}.
\end{align*}
Step (a) follows from Lemma \ref{lem:ErBound}
and (\ref{eqn:aaSSS}).
Step (b) follows from that
for any $ P_{\overline{X}_1\overline{X}_2}
         \in {\cal P}_n({\cal X}_1\times {\cal X}_2)$,
\begin{align*}
& \Psi_{\overline{X}_1\overline{X}_2}(R_1,R_2)
 2^{-nD(P_{\overline{X}_1\overline{X}_2}||P_{X_1X_2})}
\\
&=
\left\{
    2^{-n[R_1-H(\overline{X}_1|\overline{X}_2)]^{+}}
   +2^{-n[R_2-H(\overline{X}_2|\overline{X}_1)]^{+}}
   +2^{-n[R_1+R_2-H(\overline{X}_1\overline{X}_2)]^{+}}
\right\}
\cdot 2^{-nD(P_{\overline{X}_1\overline{X}_2}||P_{X_1X_2})}
\\
& \leq 3 \cdot 2^{-n[\min_{i=1,2,3} F_i(R_i|P_{X_1X_2})]}. 
\end{align*}
Step (c) follows from Lemma \ref{lem:Lem1} part a).
We next prove (\ref{eqn:mainThSecB}) in Theorem \ref{Th:mainth2}.
On an upper bound of 
$\Delta^{(n)}(\varphi^{(n)}|P_{X_1X_2}^n,$ $P_{K_1K_2}^n)$
we have the following chain of inequalities:  
\begin{align*}
& \Delta^{(n)}(\varphi^{(n)}|P_{X_1X_2}^n, P_{K_1K_2}^n)
\\
&\MLeq{a}
2(\log {\rm e})[\log(|{\cal X}_1||{\cal X}_2|)]
   n(n+1)^{2|{\cal X}_1||{\cal X}_2|}
  \sum_{ P_{\overline{K}_1\overline{K}_2}
 \in {\cal P}_n({\cal X}_1\times {\cal X}_2)}
\min\left[{1,\Theta_{\overline{K}_1\overline{K}_2}(R_1,R_2)}\right]
2^{-nD(P_{\overline{X}_1\overline{X}_2}||P_{X_1X_2})}
\\
&\MLeq{b}
6(\log {\rm e})[\log(|{\cal X}_1||{\cal X}_2|)]
n(n+1)^{2|{\cal X}_1||{\cal X}_2|}
|{\cal P}_n({\cal X}_1\times {\cal X}_2)|
2^{-n[\min\{ G_1(R_1|P_{K_1}),
             G_2(R_2|P_{K_2}),
             G_3(R_3|P_{K_1K_2})]}
\\
&\MLeq{c} 
6(\log {\rm e})[\log(|{\cal X}_1||{\cal X}_2|)]
n(n+1)^{3|{\cal X}_1||{\cal X}_2|}
 2^{-nG(R_1,R_2|P_{K_1K_2})}
=2^{-n[G(R_1,R_2|P_{K_1K_2})-\delta_{2,n}]}.
\end{align*}
Step (a) follows from Lemma \ref{lem:KeySecBound}
and (\ref{eqn:abSSS}). 
Step (b) follows from that
for any $ P_{\overline{X}_1\overline{X}_2}
         \in {\cal P}_n({\cal X}_1\times {\cal X}_2)$,
we have 
\begin{align*}
 &\min \left\{1,\Theta_{\overline{K}_1\overline{K}_2}(R_1,R_2)
       \right\}
 2^{-nD(P_{\overline{K}_1\overline{K}_2}||P_{K_1K_2})}
\\
&=\min\left\{1,
      2^{-n[H(\overline{K}_1)-R_1]}
     +2^{-n[H(\overline{K}_2)-R_2]}
     +2^{-n[H(\overline{K}_1\overline{K}_2)-(R_1+R_2)]}
    \right\} 2^{-nD(P_{\overline{K}_1\overline{K}_2}||P_{K_1K_2})}
\\
&\leq 
     \left\{ 2^{-n[H(\overline{K}_1)-R_1]^{+}}
          + 2^{-n[H(\overline{K}_2)-R_2]^{+}}
          + 2^{-n[H(\overline{K}_1\overline{K}_2)
                             -(R_1+R_2)]^{+}}
     \right\}2^{-nD(P_{\overline{K}_1\overline{K}_2}||P_{K_1K_2})}
\\
&\leq 3 \cdot 2^{-n[\min\{ G_1(R_1|P_{K_1}),
                  G_2(R_2|P_{K_2}),
                  G_3(R_3|P_{K_1K_2})\}]}.
\end{align*}
Step (c) follows from Lemma \ref{lem:Lem1} part a).
\end{IEEEproof}

\appendix


\ProofLemAA
\ProofLemA
\ProofLemB

\bibliographystyle{IEEEtran}
\bibliography{ISIT2017}

\begin{thebibliography}{1}
\providecommand{\url}[1]{#1}
\csname url@samestyle\endcsname
\providecommand{\newblock}{\relax}
\providecommand{\bibinfo}[2]{#2}
\providecommand{\BIBentrySTDinterwordspacing}{\spaceskip=0pt\relax}
\providecommand{\BIBentryALTinterwordstretchfactor}{4}
\providecommand{\BIBentryALTinterwordspacing}{\spaceskip=\fontdimen2\font plus
\BIBentryALTinterwordstretchfactor\fontdimen3\font minus
  \fontdimen4\font\relax}
\providecommand{\BIBforeignlanguage}[2]{{%
\expandafter\ifx\csname l@#1\endcsname\relax
\typeout{** WARNING: IEEEtran.bst: No hyphenation pattern has been}%
\typeout{** loaded for the language `#1'. Using the pattern for}%
\typeout{** the default language instead.}%
\else
\language=\csname l@#1\endcsname
\fi
#2}}
\providecommand{\BIBdecl}{\relax}
\BIBdecl

\bibitem{Oohama:2007:IRP:1521152.1521168}
Y.~Oohama, ``Intrinsic randomness problem in the framework of slepian-wolf
  separate coding system,'' \emph{IEICE Trans. Fundam. Electron. Commun.
  Comput. Sci.}, vol. E90-A, no.~7, pp. 1406--1417, Jul. 2007.

\bibitem{1055037}
D.~Slepian and J.~Wolf, ``Noiseless coding of correlated information sources,''
  \emph{IEEE Transactions on Information Theory}, vol.~19, no.~4, pp. 471--480,
  July 1973.

\bibitem{Csiszr1982LinearCF}
I.~Csisz{\'a}r, ``Linear codes for sources and source networks: Error
  exponents, universal coding,'' \emph{IEEE Transactions on Information
  Theory}, vol.~28, no.~4, pp. 585--592, Jul 1982.

\bibitem{KornerMarton}
J.~K{\"o}rner and K.~Marton, ``How to encode the modulo-two sum of binary
  sources (corresp.),'' \emph{IEEE Transactions on Information Theory},
  vol.~25, no.~2, pp. 219--221, Mar 1979.

\bibitem{Muramatsu03}
J.~Muramatsu, H.~Koga, and T.~Mukouchi, ``On the problem of generating mutually
  independent random sequences,'' \emph{IEICE Trans. Fundamentals}, vol. E86-A,
  no.~5, p. 1275–1284, May 2003.

\bibitem{1337277}
M.~Johnson, P.~Ishwar, V.~Prabhakaran, D.~Schonberg, and K.~Ramchandran, ``On
  compressing encrypted data,'' \emph{IEEE Transactions on Signal Processing},
  vol.~52, no.~10, pp. 2992--3006, Oct 2004.

\bibitem{DBLP:journals/corr/Oohama17b}
\BIBentryALTinterwordspacing
Y.~Oohama, ``On a relationship between the correct probability of estimation
  from correlated data and mutual information,'' \emph{CoRR}, vol.
  abs/1702.01285, 2017. [Online]. Available:
  \url{http://arxiv.org/abs/1702.01285}
\BIBentrySTDinterwordspacing

\bibitem{csi2011information}
\BIBentryALTinterwordspacing
I.~Csisz{\'a}r and J.~K{\"o}rner, \emph{Information Theory: Coding Theorems for
  Discrete Memoryless Systems}.\hskip 1em plus 0.5em minus 0.4em\relax
  Cambridge University Press, 2011. [Online]. Available:
  \url{https://books.google.co.jp/books?id=2gsLkQlb8JAC}
\BIBentrySTDinterwordspacing

\bibitem{DBLP:journals/tit/OohamaH94}
\BIBentryALTinterwordspacing
Y.~Oohama and T.~S. Han, ``Universal coding for the slepian-wolf data
  compression system and the strong converse theorem,'' \emph{{IEEE} Trans.
  Information Theory}, vol.~40, no.~6, pp. 1908--1919, 1994. [Online].
  Available: \url{http://dx.doi.org/10.1109/18.340465}
\BIBentrySTDinterwordspacing

\end{thebibliography}

\end{document}